\documentclass[twocolumn,twocolappendix,tighten]{aastex63}
\usepackage{graphicx}
\usepackage{url}
\usepackage{epstopdf}
\usepackage{gensymb}
\usepackage{multirow}
\usepackage{ragged2e}
\usepackage{hyperref}
\usepackage{url}
\usepackage{amsmath} 
\usepackage{booktabs}
\usepackage{comment}
\usepackage{afterpage}
\usepackage{mathtools}
\usepackage{tabularx}
\usepackage{bold-extra}
\usepackage{xspace}
\usepackage{relsize}
\usepackage{braket}
\usepackage[nolist,nohyperlinks]{acronym} 
\usepackage{eht}
\usepackage[T1]{fontenc} 
\usepackage[encapsulated]{CJK}
\usepackage{ucs}
\newcommand{\cntext}[1]{\begin{CJK}{UTF8}{gbsn}#1\end{CJK}}
\usepackage{apjfonts}
\usepackage{textcomp} %PT adds
\usepackage{color}
\usepackage{mathtools}

\newcommand{\TBD}[1]{\textit{\textcolor{red}{FLAG}}}

\def\sgra{\object{Sgr~A$^{\ast}$}\xspace}
\def\m87{\object{M87$^{\ast}$}\xspace}
\def\bllac{\object{BL~Lac}\xspace}
\def\cta{\object{CTA 102}\xspace}
\def\jfour{\object[0420-014]{J0423$-$0120}\xspace}
\def\jfiveone{\object[0507+179]{J0510+1800}\xspace}
\def\jfivetwoone{\object{J0521+1638}\xspace}
\def\jfivetwo{\object[0521-365]{J0522$-$3627}\xspace}
\def\threec{\object{3C~454.3}\xspace}
\def\um{~\object{\textmu m}\xspace}
\def\uas{~\object{\textmu as}\xspace}

\begin{document}
\title{First Very Long Baseline Interferometry Detections at 870\um}
\shorttitle{First $870\!\um$ VLBI}
\shortauthors{Raymond \& Doeleman et al.}
\author[0000-0002-5779-4767]{Alexander W. Raymond}
\affiliation{Black Hole Initiative at Harvard University, 20 Garden Street, Cambridge, MA 02138, USA}
\affiliation{Center for Astrophysics $|$ Harvard \& Smithsonian, 60 Garden Street, Cambridge, MA 02138, USA}
\affiliation{Current address is Jet Propulsion Laboratory, California Institute of Technology, Pasadena, CA, USA}

%\correspondingauthor{Author Name}
\author[0000-0002-9031-0904]{Sheperd S. Doeleman} 
\altaffiliation{Corresponding Author: S. S. Doeleman, sdoeleman@cfa.harvard.edu}
\affiliation{Black Hole Initiative at Harvard University, 20 Garden Street, Cambridge, MA 02138, USA}
\affiliation{Center for Astrophysics $|$ Harvard \& Smithsonian, 60 Garden Street, Cambridge, MA 02138, USA}

\author[0000-0001-6988-8763]{Keiichi Asada}
\affiliation{Institute of Astronomy and Astrophysics, Academia Sinica, 11F of Astronomy-Mathematics Building, AS/NTU No. 1, Sec. 4, Roosevelt Rd., Taipei 106216, Taiwan, R.O.C.}

\author[0000-0002-9030-642X]{Lindy Blackburn}
\affiliation{Black Hole Initiative at Harvard University, 20 Garden Street, Cambridge, MA 02138, USA}
\affiliation{Center for Astrophysics $|$ Harvard \& Smithsonian, 60 Garden Street, Cambridge, MA 02138, USA}

\author[0000-0003-4056-9982]{Geoffrey C. Bower}
\affiliation{Institute of Astronomy and Astrophysics, Academia Sinica, 
645 N. A'ohoku Place, Hilo, HI 96720, USA}
\affiliation{Department of Physics and Astronomy, University of Hawaii at Manoa, 2505 Correa Road, Honolulu, HI 96822, USA}

\author{Michael Bremer}
\affiliation{Institut de Radioastronomie Millim\'etrique, 300 rue de la Piscine, F-38406 Saint Martin d'H\`eres, France}

\author[0000-0001-9151-6683]{Dominique Broguiere}
\affiliation{Institut de Radioastronomie Millim\'etrique, 300 rue de la Piscine, F-38406 Saint Martin d'H\`eres, France}

\author[0000-0001-6573-3318]{Ming-Tang Chen}
\affiliation{Institute of Astronomy and Astrophysics, Academia Sinica, 11F of Astronomy-Mathematics Building, AS/NTU No. 1, Sec. 4, Roosevelt Rd., Taipei 106216, Taiwan, R.O.C.}

\author[0000-0002-2079-3189]{Geoffrey B. Crew}
\affiliation{Massachusetts Institute of Technology Haystack Observatory, 99 Millstone Road, Westford, MA 01886, USA}

\author{Sven Dornbusch}
\affiliation{Max-Planck-Institut f\"ur Radioastronomie, Auf dem H\"ugel 69, D-53121 Bonn, Germany}

% ...

\begin{abstract}
The first very long baseline interferometry (VLBI) detections at 870\um wavelength (345~GHz frequency) are reported, achieving the highest diffraction-limited angular resolution yet obtained from the surface of the Earth, and the highest-frequency example of the VLBI technique to date.  These include strong detections for multiple sources observed on inter-continental baselines between telescopes in Chile, Hawaii, and Spain, obtained during observations in October 2018.  The longest-baseline detections approach 11~G$\lambda$ corresponding to an angular resolution, or fringe spacing, of 19\uas.  The Allan deviation of the visibility phase at 870\um is comparable to that at 1.3~mm on the relevant integration time scales between 2 and 100~s.  The detections confirm that the sensitivity and signal chain stability of stations in the Event Horizon Telescope (EHT) array are suitable for VLBI observations at 870\um.  Operation at this short wavelength, combined with anticipated enhancements of the EHT, will lead to a unique high angular resolution instrument for black hole studies, capable of resolving the event horizons of supermassive black holes in both space and time. 

\end{abstract}

\section{Introduction}
\begin{figure}[h!]
\centering
  \includegraphics[width=0.9\columnwidth]{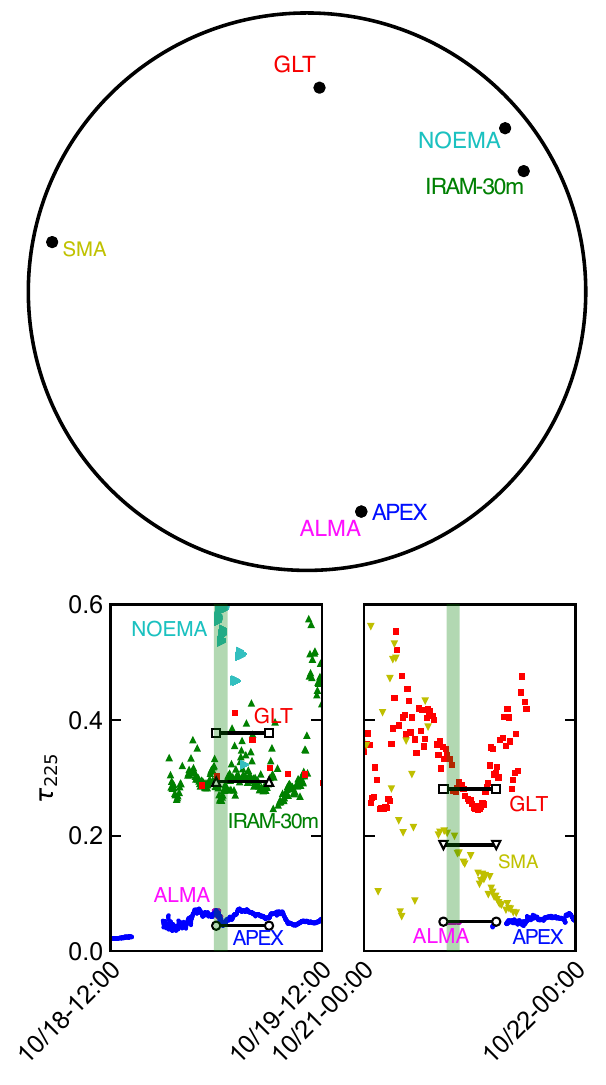}
  \caption{(top) Stations in the 870\um fringe test.  (bottom) Zenith opacity at 225~GHz, which is the standard frequency used for monitoring mm-wave conditions.  The observing window on each day is indicated by the green shading.  Conditions at ALMA were very good during both days ($\tau_{225} \approx 0.05$).  The black lines indicate the opacity at each site calculated using inputs from MERRA-2 reanalysis during the observing windows, which we use to estimate 870\um (345~GHz) opacity.  Opacities for APEX and NOEMA have been estimated by converting precipitable water vapor column amounts.}
  \label{fig:experiment}
\end{figure}

The technique of very long baseline interferometry (VLBI) involves a network of independently clocked telescopes separated by large distances, which simultaneously observe a common astronomical source~\citep{TMS2017}.  The angular resolution, or fringe spacing, in a VLBI observation scales inversely with both the distance between stations (\textit{i.e.}, the length of the baseline) and the observing frequency.  The present article reports the first fringe detections made at 870\um wavelength (345~GHz nominal frequency),  which constitutes the shortest wavelength VLBI observation to date.  The experiment we describe was intended as a first technical demonstration of the 870\um VLBI capability using facilities that are part of the Event Horizon Telescope (EHT) array.  Figure~\ref{fig:experiment} shows the stations that participated in the fringe test along with the usual metric used to characterize mm-wavelength observing conditions: the 225~GHz zenith opacity \citep{TMS2017}.

VLBI observing wavelength has decreased over time.  The first 3~mm VLBI detections (at 86~GHz) were obtained through observations performed in 1981~\citep{readhead1983}; the first 3~mm intercontinental detections (100~GHz) were obtained through observations performed in 1988~\citep{baath1991,baath1992}, and the first successful 1.3~mm (230~GHz) VLBI was carried out in 1989~\citep{padin1990}.  The especially long time since the last significant decrease in VLBI wavelength reflects the challenges of carrying out such observations, which are detailed below. Even so, there have been several milestones of note since the early 1990s on the path towards developing short wavelength VLBI as an important technique for astrophysics.  Increased sensitivity through the use of larger telescopes and advanced receivers led to 1.4~mm (215~GHz)  detections on a $\sim$1100~km baseline of multiple active galactic nuclei (AGN) and Sagittarius~A* (Sgr~A*), the Galactic Center supermassive black hole \citep{Greve_1995,Krichbaum_1997,Krichbaum_1998}. A return to the longer-wavelength 2~mm spectral windows (147~GHz and 129~GHz) allowed extension of mm-wavelength VLBI to intercontinental baselines \citep{Greve_2002,Krichbaum_2002,Doeleman_2002}.  Building on this work, \citet{doeleman2008,Doeleman2012} used purpose-built wideband digital VLBI systems on 1.3~mm trans-oceanic baselines to report the discovery of event-horizon scale structures in \sgra and the much more massive black hole, \m87.  The Event Horizon Telescope (EHT) collaboration has now imaged both of these sources with a global 1.3~mm VLBI array \citep{paperi,sgr_paper1,M87_2018_aa}.

%=========FIGURE 1===========
%============================

The EHT is the highest-resolution ground-based VLBI instrument to date \citep{paperii}.  The EHT fringe spacing is approximately 25\uas at 1.3~mm wavelength.  The finite diameter of the Earth limits ground-based 1.3~mm fringe spacing to 21\uas corresponding to 9.8 G$\lambda$ baseline. In practice, modern imaging methods, such as regularized maximum likelihood, achieve a slightly higher angular resolution that exceeds the diffraction limit~\citep{paperiv}.
  
For future campaigns, the EHT has developed the capability to observe at 870\um, and enhancing the ability to observe at this wavelength through new stations and wider bandwidth is an important aspect of long-term enhancements envisaged by the next-generation EHT (ngEHT) project \citep{Doeleman_2019, raymond2021,ngeht_refarray}.  For a given set of station locations, observing at 870\um improves angular resolution by approximately 50\% compared to observing at 1.3~mm, which will provide a sharper view of the black hole shadow and environment; the 870\um fringe spacing limit set by the diameter of the Earth is approximately 14\uas corresponding to 14.7~G$\lambda$ baseline.  Observations at 870\um are also important for polarimetric measurements.  Faraday rotation, which scrambles the imaged electric field vector position angle pattern, diminishes with the square of the frequency.  Therefore, 870\um observations may help distinguish Faraday rotation from the intrinsic field pattern set by the horizon-scale magnetic field and plasma properties~\citep{paperviii,Wielgus_scattering_screen_2023}.  For \sgra, the angular size of the black hole shadow is larger than that of \m87 \citep{sgr_paper1}, but scattering in the ionized interstellar medium affects the image angular resolution (see, e.g., \citealt{johnson2018}). At 1.3~mm, the scatter-broadening is comparable to the current EHT resolution, but it decreases approximately as the observing wavelength squared. Thus, at 870\um, scattering effects would be significantly diminished and would not limit the resolution of a VLBI array for studies of \sgra.  In particular, extension of the EHT to 870\um wavelengths can target photon ring substructure in \sgra, aiming to detect the orbit of light that makes a full ``u-turn" around the black hole \citep{Johnson_2020_photon_ring, Palumbo_2023}.  For these reasons, 870\um VLBI opens important new directions for advanced horizon-resolved studies of the two primary EHT sources.  At the same time, higher frequency VLBI brings more sources into range for horizon-resolved black hole studies \citep{Pesce_2021,Venki_2023,Lo_2023}, and the increased resolution at 870\um benefits non-horizon VLBI studies of active galactic nuclei (AGN) jets  \cite[e.g.,][]{kim2020,janssen2021,Issaoun2022,Jorstad2023,Paraschos2024}.  
Additionally, due to reduced opacity, shorter wavelengths probe more compact regions of jetted AGN sources \citep[an example being the core-shift effect:][]{lobanov1998,Hada_2011}. Hence, 870\um VLBI has the potential to image the jet launching region closer to the central black hole, enabling investigations of the physics behind jet formation, collimation, and acceleration. In particular, the poorly understood limb-brightening in transversely resolved inner jets (e.g. \citealt{janssen2021}) can be studied in much greater detail.

Extension of observing to 870\um similarly enhances the capability of the EHT to capture dynamics near the event horizon.  In the case of \sgra, the dynamical time scale is $\sim200$s ($10 GM/c^3)$.  Simultaneous 1.3~mm and 870\um observing can sample sufficient Fourier spatial frequencies within this integration time to allow snapshot imaging using the technique of multi-frequency synthesis \citep[MFS;][]{Chael_MFS_2023}.  Combining such snapshots will enable recovery of accretion and jet launching kinematics.  For \m87, the dynamical time scale is $\sim$3 days, and data obtained in both 1.3~mm and 870\um on sequential days can be combined to form high-fidelity MFS images for time-lapse movie reconstruction of the event horizon environment.  Realizing the full scientific potential of 870\um VLBI \citep{ngeht_ksg} will require the planned ngEHT upgrade \citep{ngeht_refarray}.

While there are clearly many motivating reasons for 870\um VLBI observing, a number of factors make the measurements difficult in this short-wavelength regime.  The atmosphere is more opaque at 870\um than at 1.3~mm (see for example \cite{liebe1985,Matsushita_1999,Matsushita_2017,Matsushita_2022}), which means that sources are more attenuated and noise levels due to atmospheric emission are elevated.  Overall, the effective system temperatures of coherent radio receivers are intrinsically greater at 870\um than at 1.3~mm\footnote{See, for example, \citet{janssen2019} or ALMA Cycle 8 2021 Technical Handbook.}.  The aperture efficiency of the collecting optics tends to diminish at high frequency, and the source flux density tends to decrease.  In addition, coherence losses due to the VLBI frequency standards used at each site increase with observing frequency \citep{Doeleman_2011}.  The EHT array, conceived as a common, international effort of independent observatories working in the short millimeter range, has directly addressed these challenges and provides key enabling infrastructure for extension of VLBI to higher frequencies~\citep{paperii}. 

The telescopes comprising the EHT array are precision structures sited at high-altitude, low-opacity locations (see e.g. \cite{levy1996structural}, \cite{2006PASP..118.1257M}, \cite{2010ASSL..364.....G}, \cite{Chen_2023} and references therein on the design and qualification of such instruments).  State of the art instrumentation underpinning the operation of these telescopes, as single-dish facilities and for VLBI, includes cryogenic receivers and wideband digital backends - all refined over many years to optimize performance at mm and submm wavelengths. Steady improvements in superconductor-insulator-superconductor (SIS) junctions have formed the basis for increased bandwidth and sensitivity of mm and submm receivers, leading to state-of-the-art systems in use at EHT sites (see \cite{maier2005}, \cite{tong2005}, \cite{chenu2007}, \cite{carter2012}, \cite{maier2012}, \cite{mahieu2012}, \cite{Tong_2013}, \cite{Kerr_2014}, \cite{chenu2016}, \cite{klein2014}, \cite{chihchiang2018}, \cite{Belitsky_2018}).

Following the successful 1.3~mm VLBI observations in 2017, test observations at 870\um were conducted on the EHT array in October 2018.  Conditions at the ALMA station during this test, including characterization of the system used there to phase the array for VLBI, are described in ~\cite{crew2023}.  The present paper describes the VLBI test observations 
\section{Methods}\label{sec:methods}
\subsection{Schedule}
The 870\um fringe test observations consisted of two short scheduling blocks designed for two different subarrays.  An eastern subarray, comprising ALMA, the Atacama Pathfinder EXperiment (APEX), Greenland Telescope (GLT), the Institut de Radioastronomie Millim\'etrique 30~m telescope (IRAM30m), and the Northern Extended Millimeter Array (NOEMA), was scheduled to include blazar sources that were visible in the nighttime hours at all sites: \cta, \threec, and \bllac.  A western subarray, comprising ALMA, APEX, GLT, and the Submillimeter Array (SMA), observed quasars \jfour, \jfiveone, \jfivetwoone, and \jfivetwo.   The eastern subarray scheduling block was followed by several scans on \bllac at 1.3~mm wavelength to aid diagnosis in the event of a null result.   Schedule blocks for both subarrays were optimized for fringe detection at 870\um VLBI, and they spanned a duration of between 1 and 2 hours with at least two scans on every source.  Most scans lasted five minutes.

\begin{figure}[h!]
\centering
  \includegraphics[width=1.\columnwidth]{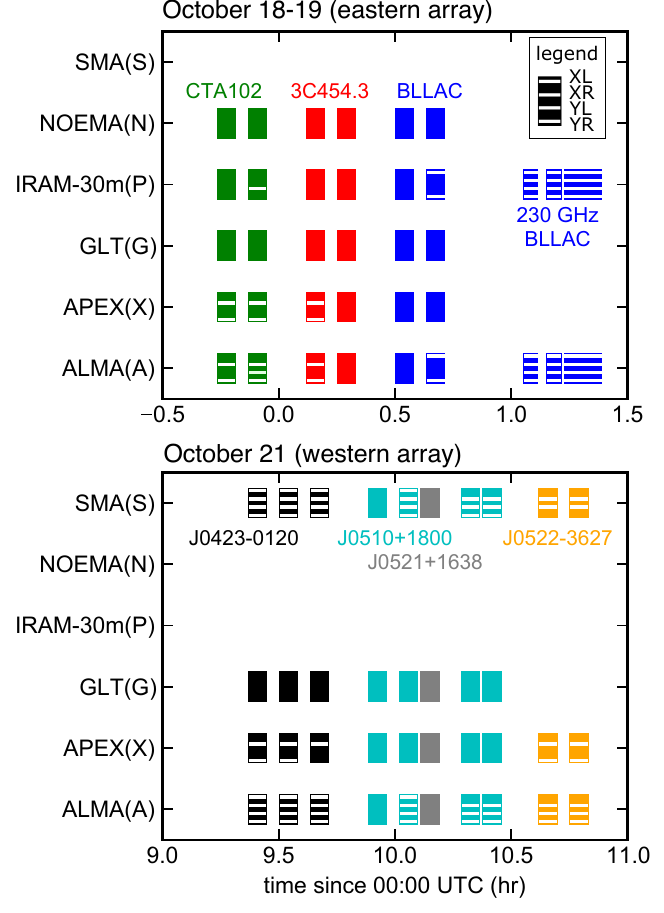}
  \caption{870\um observations that yielded detections were made during two separate scheduling blocks: October 18/19 and October 21, 2018.  The observations on the first night were done with an eastern array comprising ALMA, APEX, GLT, IRAM30m, and NOEMA.  Observations on the second night were made with a western array: ALMA, APEX, GLT, and SMA.  The scheduling blocks for both nights are shown along with the one-letter station codes, which are listed in parenthesis.  All detections are on baselines involving ALMA.  The scans which yielded detections on baselines defined by a given station are indicated by the white horizontal ticks centered in each time block: from the top, ticks correspond to XL, XR, YL, YR mixed-polarizations per the legend shown upper right.  The absence of a tick indicates a non-detection.  Three scans at 230~GHz (1.3~mm) were performed at the end of the eastern subarray scheduling block using just the IRAM30m and ALMA facilities.}
  \label{fig:schedule}
\end{figure}

The observing window consisted of five nights 2018 October 17--21 between approximately midnight and 2:00 Coordinated Universal Time (UTC) for the eastern subarray scheduling block and between 9:00 and 11:00 UTC for the western subarray scheduling block.  Each scheduling block was triggered twice within the observing window.  We report herein on successful observations with the eastern array on 2018 October 18-19 and with the western array on 2018 October 21. Details of the scheduling blocks and sources observed are shown in Fig.~\ref{fig:schedule}.

%=========FIGURE 2===========
%============================

\subsection{Instrumentation and Array}\label{sec:inst}

Several important technologies developed for 1.3\,mm VLBI are leveraged to address the challenges of 870\um  observing, many of which are outlined in \citet{paperii}.  The VLBI backends, used to condition and digitize signals from the telescope receivers, have a cumulative data rate of 64~Gbps~\citep{Vertatschitsch2015,Tuccari2017} across four 2-GHz wide bands and two polarizations. Each station is outfitted with a hydrogen maser time standard, which had previously been found to be sufficiently stable for timekeeping in a 1.3\,mm VLBI experiment and were expected to be sufficiently stable for 870\um. 

Phased array beamforming capability is implemented at both the SMA~\citep{young2016} and ALMA~\citep{matthews2018} array stations.   For both these stations beamformer phasing efficiency at 870\um, which directly scales the visibility amplitudes measured on baselines to the station, varied from just below 50\% to as high as about 80\%.  These efficiencies are less than what is typical for 1.3~mm~\citep{paperii}.  Section \ref{sec:phasing} has discussion  relevant to ALMA, SMA, and also to NOEMA\footnote{NOEMA is also equipped with the phased array though it was not commissioned at the time of this observation.} of phasing efficiency challenges and planned improvements to mitigate these.

The frequency setup for the  870\um fringe test is similar to that described in Table~4 of \citet{paperii}.  Most stations in the array observed a single 2048~MHz band at a 4--6~GHz intermediate frequency (IF) using a 342.6~GHz sky local oscillator (LO)\footnote{ALMA and SMA used slightly different frequency setups to match the sky frequency of the other stations, see sections \ref{sec:alma} and \ref{sec:sma}.}.  That frequency setup corresponds to a sky frequency range of 346.552 to 348.6~GHz.  Each station observed both circular polarizations, with the exceptions of APEX (right-circular polarization, RCP, only) and ALMA (dual linear, X and Y).   The recorded station data were correlated using  DiFX software  \citep{deller2011} at the MIT Haystack Observatory. Visibility data on baselines to ALMA remained in a mixed-polarization basis (\textit{i.e.}, \{X,Y\} $\times$ \{L,R\}) because the observing schedules were not long enough to track polarization calibrators over a wide range of parallactic angle, which is necessary for converting the ALMA data from a linear to circular basis~\citep{martividal2016,matthews2018,goddi2019}.  Subsequent fringe fitting was done using the Haystack Observatory Post-processing System (HOPS\footnote{\url{https://www.haystack.mit.edu/tech/vlbi/hops.html}}, \cite{whitney2004}; see also \cite{blackburn2019a}).

\subsubsection{ALMA}\label{sec:alma}
ALMA observed in dual linear polarization with  IRAM designed 870\um (\textit{i.e.}, Band 7) cartridges~\citep{mahieu2012}. 
 The ALMA Phasing System (APS)~\citep{matthews2018} was used to aggregate the collecting area of the active dishes in the ALMA array.  The APS capability had been used previously for VLBI science at 3~mm~\citep{issaoun2019,okino2022,zhao2022} and 1.3~mm~\citep{paperi, paperii} but not at shorter wavelengths albeit that setup for 870\um observations is similar to the longer wavelength bands.  In the 870\um experiment, the four recorded 2.048\,GHz subbands were tuned to center frequencies of 335.6, 337.541406, 347.6 and 349.6 GHz.  The choice of the 337.541406 GHz frequency results from ALMA-specific tuning restrictions.  

The ALMA phased array included twenty-five 12~m antennas during the eastern track and twenty-nine 12~m antennas during the western track with a maximum antenna spacing of 600~m in both cases.  Wind speeds were greater than 10~m~s\textsuperscript{$-1$} at the ALMA site.  During the Eastern track, phasing efficiency was below 50\% for most of the time and at best was about 80\%.  During the October 21 track (western) in better weather, phasing efficiency was more stable and greater than approximately 90\%~\citep{crew2023}.  

\subsubsection{APEX} \label{sec:apex}
The APEX and ALMA stations are co-located and conditions were similar at the two telescopes.  APEX observed using the 345\,GHz FLASH+ linear receiver \citep{Klein_2014}.   That receiver may not have been functioning optimally during the experiment and has since been replaced by the  Swedish-ESO PI Instrument for APEX (SEPIA) \citep{Belitsky_2018,meledin_2022}. A quarter wave plate was used to achieve circular polarization. Two backends, a ROACH2 Digital Backend ~\citep[R2DBE;][]{Vertatschitsch2015} and a Digital BaseBand Converter 3 ~\citep[DBBC3;][]{Tuccari2017}, were operated in parallel.

\subsubsection{GLT}
The GLT station participated in the observation but at the time was still commissioning specific subsystems.  The GLT antenna has operated at Pituffik Space Base, formerly the Thule Airbase site, in Greenland since August 2017~\citep{Inoue_2014,raffin2016,matsushita2018,Koay_2020,Chen_2023}.  The GLT observed in dual linear polarization with the IRAM-made 870\um (\textit{i.e.}, Band 7) cartridges~\citep{mahieu2012}. The 345 GHz receiver on the GLT saw first-light in continuum and spectral-line modes in August 2018. Pointing and focus calibration at 345~GHz were still in the commissioning phase during the 870\um observation reported here. The GLT pointing system has since been fully commissioned for recent and future VLBI observing.  Similarly, final adjustments to the dish surface had yet to be made, and the surface accuracy was estimated to be 170$\mu$m rms during the observations reported here. Subsequent improvements have led to rms surface accuracy in the 17-40$\mu$m range (see Table 7 in \citet{Chen_2023}).

\subsubsection{IRAM30m}
The IRAM30m telescope used the heterodyne Eight MIxer Receiver \citep{carter2012} in the 870\um band also known as E330.  The setup and pre-observing checks were analogous to a regular Global Millimeter VLBI Array or EHT session.  The opacity at 870\um during the scheduled VLBI observations was high and would not typically have triggered single-dish science operation at this wavelength.

\subsubsection{NOEMA}
Portions of the NOEMA station were still being commissioned during the 870\um experiment.  NOEMA observed in dual polarization as a single-antenna station not as a phased array.  The NOEMA receiver was a dual-polarization single-sideband unit~\citep{chenu2016} with a 4 GHz bandpass.  Recording was with a 16~Gbps R2DBE backend.  The NOEMA phased array has since been commissioned for VLBI observing.

\subsubsection{SMA}\label{sec:sma}
The SMA station observed with seven antennas arranged in the compact configuration with a maximum baseline of 69.1~m.  The SMA Wideband Astronomical ROACH2 Machine (SWARM)~\citep{primiani2016,young2016} was run with the VLBI beamformer mode activated producing a coherent phased array sum of the seven antennas, formatted for VLBI recording. As expected the phasing efficiency was lower than for 1.3\,mm operations.   The sky LO was set to 341.6~GHz, not 342.6~GHz, to match the SWARM sky coverage with the other stations, compensating for a different IF to baseband local oscillator because SWARM uses its own block downconverter rather than the standard EHT single dish equipment.  The data were recorded in the frequency domain at the standard SMA clock rate (4.576 Gsps) which differs from the standard EHT single dish sample rate of 4.096 Gsps \citep{Vertatschitsch2015}. APHIDS (Adaptive Phased Array Interpolating Downsampler for SWARM) post-processing was completed to interpolate and invert (from frequency- to time-domain) the SWARM data sets in preparation for VLBI correlation.  After APHIDS processing the SMA EHT data product matches that produced by standard SMA single dish station in sample rate, and is also a time series matching the standard EHT single dish data product.  

\section{Results and Discussion}
Figure~\ref{fig:experiment} shows that the conditions during the experiment were mixed across the array.  While the observatories do not measure 870\um (345~GHz) opacity directly, we use MERRA-2 reanalysis and radiative transfer~\citep{paine2022} that is validated by measurements at 225~GHz (Fig.~\ref{fig:experiment} black lines) to estimate $\tau_{345}$. For the eastern subarray on October 18/19, $\tau_{345}$ was 0.2 at the ALMA and APEX sites, and 0.8 at IRAM30m.  For the western subarray on October 21, $\tau_{345}$ was approximately 0.17 at the ALMA and APEX sites and 0.7 at SMA.  During the experiment, the opacities at GLT and NOEMA were unfavorable and detections on baselines to those stations were not achieved; however, both stations have weather that is compatible with 870\um observing and will likely yield high-frequency detections in the future (see e.g., \citet{raymond2021, Matsushita_2022}).  Atmospheric conditions can change rapidly: $\tau_{225}$ at the SMA decreased by nearly a factor of four in the hours following the experiment.

\subsection{$870\mu$m (345~GHz) Fringes}
\begin{figure}[h]
\centering
  \includegraphics[width=1\columnwidth]{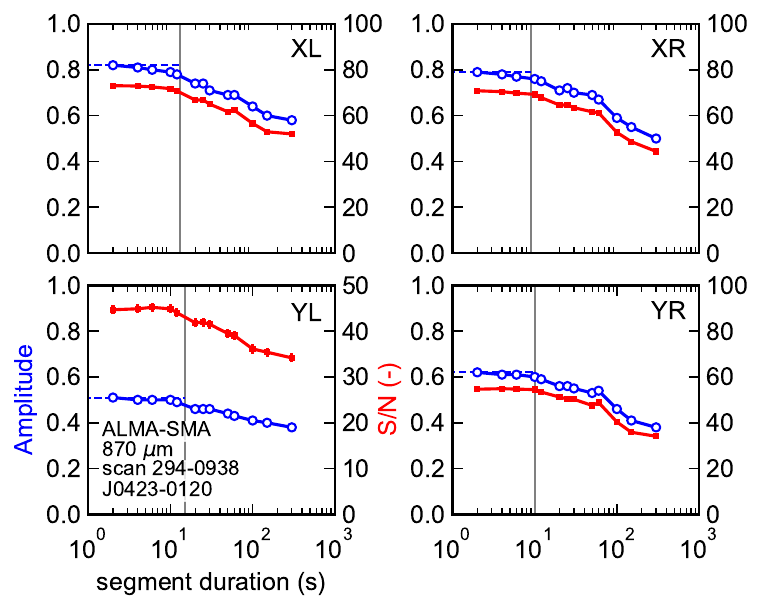}
  \caption{Scan averaged and noise-debiased 870\um fringe amplitude (open blue circles, left axes) and $S/N$ (closed red squares, right axes).  Amplitudes and $S/N$ are computed by first dividing each observing scan into short coherently integrated segments, which are then combined incoherently following the procedure in \citet{rogers1995}. Segment length is shown on the horizontal axis.  Each subplot shows a different polarization on the ALMA-SMA baseline for a single scan on \jfour (October 21, 09:38 UTC).  Other detections listed in Table~\ref{table:experiments345} have similar dependence on segment duration though generally lower $S/N$.  The noise-debiased amplitude and coherence time were derived using HOPS and are indicated by the horizontal blue dashed line and the vertical solid black line, respectively.}
  \label{fig:coherencetime}
\end{figure}

In VLBI, recorded data from all sites are brought to a central processing facility where data streams from each pair of sites are cross-correlated.  The resulting complex correlation quantities provide a dimensionless measure of the electric field coherence between the two sites, which is proportional to a Fourier component of the brightness distribution of the target source.  The correlation processor uses an apriori model to align the site data streams, recreating the exact geometry of the physical baseline connecting the two sites at the time of observation.    
Because the apriori model is imperfect, after processing the cross correlation phase typically varies as a function of time and frequency due to residual delay and delay-rate respectively. To average the correlation signal over frequency and time, the correlator output is thus searched over a range of delay and delay-rate to find a peak in correlator power - a process also known as 'fringe-fitting' \citep{TMS2017}.     In this experiment, the correlator output was searched by dividing each scan into short segments and incoherently averaging them.  The incoherent averaging technique \citep{rogers1995} estimates noise-debiased VLBI quantities, and it is well suited to processing low-$S/N$ VLBI data on sparse arrays as it allows integration beyond the nominal atmospheric coherence time. Figure~\ref{fig:coherencetime} shows the dependence of amplitude  in units of 10$^4$ and signal-to-noise ratio ($S/N$) on the duration of the segments for a sample scan on source \jfour for the baseline comprising the ALMA and SMA stations.  All four cross-hand polarizations are plotted.  The scan identifier 294-0938 in Fig.~\ref{fig:coherencetime} corresponds to the \textit{day}-\textit{UTC} for the beginning of the scan, where the \textit{day} is the number of days since January 1, 2018 (294 is October 21) and \textit{UTC} is the scan start time.  The noise-debiased amplitude~\citep{rogers1995} in Fig.~\ref{fig:coherencetime} is indicated by the dashed horizontal line.  As the segment duration decreases, the effect of decoherence is reduced so the $S/N$ increases.

Compared to a single coherent integration over a full scan (approximately 300~s in most of the measurements), incoherently averaging the parts of a segmented scan increases the $S/N$ by up to a factor of two on many of the measurements, yielding higher confidence in the detections.  For most of the measurements, $S/N$ values asymptote at the shortest segment durations.  Ordinarily, we would expect the $S/N$ values to decrease as the segments are shortened below the coherence time.  The behavior we observe could be indicative of a changing coherence during the scan consistent with the windy conditions at ALMA \citep{crew2023}.

%=========FIGURE 3===========
%============================

\begin{figure}[h!]
\centering
  \includegraphics[width=1\columnwidth]{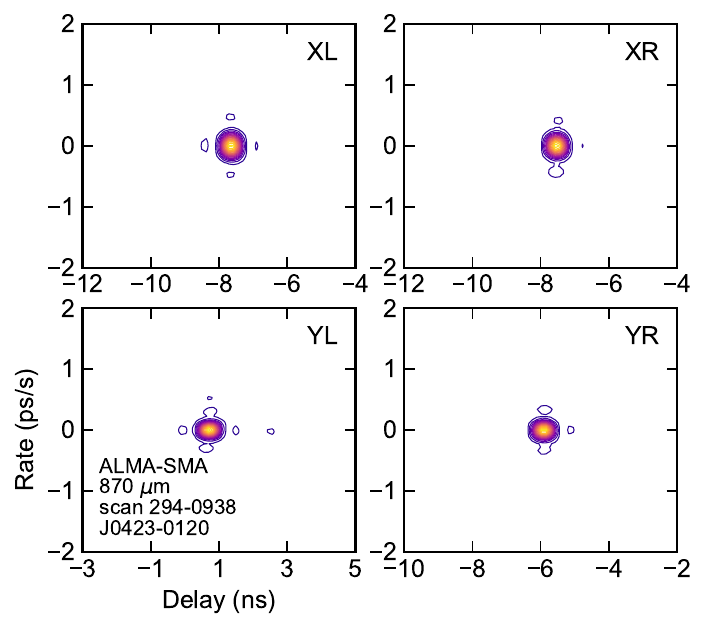}
  \caption{870\um contours of incoherently-averaged fringe power in 5\% increments versus delay and rate for a single scan on \jfour for the ALMA-SMA baseline (October 21, 09:38 UTC).  Other detections reported in Table~\ref{table:experiments345} also exhibit clear peaks versus delay/rate.}
  \label{fig:delayrate}
\end{figure}

Contours of fringe power versus multi-band delay and rate are plotted in Fig.~\ref{fig:delayrate} for a single scan of \jfour on the ALMA-SMA baseline.  The measurement exhibits a definitive peak in fringe power for each of the cross-hand polarizations.  The rates are all centered near zero.
Multi-band delays fall within an ambiguity search window of (-8.53~ns, 8.53~ns) as they are derived from measurements spaced at ALMA's channel separation of 58.592375~MHz \citet{matthews2018,paperiii}).

%=========FIGURE 4===========
%============================

The fringe detection threshold was conservatively set at $S/N$>7 to prevent false detections, and all resulting detections are summarized in Table~\ref{table:experiments345} ordered by target source.  The maximum spatial frequencies sampled are greater than 10.9~G$\lambda$ between ALMA and the SMA, which significantly exceeds the largest spatial frequencies sampled by the EHT for \m87 at 1.3~mm on the longest baseline between Hawaii and Europe (approximately 8~G$\lambda$).  The highest $S/N$ detections exceed 70.  Simultaneous detections in all four polarization products were achieved on the ALMA-SMA baseline for \jfour.  The zero-baseline flux densities at 870\um were obtained from the ALMA local interferometry~\citep{crew2023}.  The flux densities were 1.4, 1.0, 2.4, 1.2, and 4.9~Jy on \cta, \bllac, \jfour, \jfiveone, and \jfivetwo, respectively.  The source structure of the targets in this work is not known apriori, so it is not possible to say with precision how the correlated amplitudes should vary as a function of baseline length.  Furthermore, these observations were designed to be a detection experiment, and not carried out with all procedures that would allow robust VLBI flux density calibration.  Nevertheless, the SNR on the ALMA-APEX baselines appears to be anomalously low given the short baseline length, which would ordinarily be sensitive to both small scale structure (10-100$\mu$as) and larger scale structure (10-100mas).  This is likely attributable to phase instabilities suspected in the APEX receiver (see Section~\ref{sec:apex}), which has since been retired.  Follow-on experiments, already scheduled, will focus on calibration and robust flux density measurements vs. baseline length.

HOPS reports two coherence times: one corresponding to the point below which there is only a small amount of coherence loss within the uncertainty of amplitudes and another corresponding to the maximum $S/N$.  For most of the scans in Table~\ref{table:experiments345}, we report the former.  In a few low-$S/N$ cases where the routine was unable to fit the coherence, the coherence time based on $S/N$ is reported instead.  The coherence times across baselines range from approximately ten to thirty seconds for most cases.  For \bllac, the longer coherence times may be an artifact of the moderate $S/N$.

%=========TABLE 1===========
\begin{deluxetable*}{ccccccccccccc}
\tablecaption{870\um detections on the indicated baselines, sources, and polarizations.\label{table:experiments345}}
\tabletypesize{\footnotesize}
\tablehead{
\colhead{\textbf{Baseline$^\dag$}} &
\colhead{\textbf{Pol.}} &
\colhead{\textbf{Day$^*$}} &
\colhead{\textbf{Time} (hh:ss)} &
\colhead{\textbf{El. 1 (\degr)}} &
\colhead{\textbf{El. 2 (\degr)}} &
 \colhead{$\mathbf{|\vec{u}}$-$\mathbf{\vec{v}|}$ (G$\lambda$)} &
\colhead{\textbf{$\tau_c$} (s)} &
 \colhead{\textbf{Delay} (ns)} &
 \colhead{\textbf{Rate} (fs~s$^{-1}$)} &
\colhead{\textbf{Amp.} ($\times10^{-4}$)} &
\colhead{\textbf{S/N}} 
}
\startdata
\multicolumn{11}{c}{\threec}\\
\hline
AX&XR&292&00:07&44.9&45.0&0.0026&8&4.4&-1&0.50&43.7\\
AX&YR&292&00:07&44.9&45.0&0.0026&8&5.2&-1&0.47&41.4\\
\hline
\multicolumn{11}{c}{\bllac}\\
\hline
AP&XL&292&00:38&24.6&42.6&9.7913&31&-4.6&4&0.15&12.2\\
AP&YR&292&00:38&24.6&42.6&9.7913&46&-8.5&0&0.13&10.8\\
\hline
\multicolumn{11}{c}{\cta}\\
\hline
AP&YL&291&23:52&49.7&43.5&9.9581&21&0.9&-38&0.18&13.6\\
AX&XR&291&23:44&48.6&48.7&0.0027&24&5.6&-38&0.23&19.2\\
AX&XR&291&23:52&49.7&49.7&0.0027&10&5.2&-85&0.23&20.8\\
AX&YR&291&23:44&48.6&48.7&0.0027&22&6.3&-51&0.21&17.6\\
AX&YR&291&23:52&49.7&49.7&0.0027&11&6.0&-84&0.22&18.0\\
\hline
\multicolumn{11}{c}{\jfour}\\
\hline
AS&XL&294&09:22&48.5&35.5&10.8547&14&-7.6&6&0.54&47.8\\
AS&XL&294&09:30&46.8&37.3&10.8874&14&-8.0&0&0.70&62.4\\
AS&XL&294&09:38&45.1&39.1&10.9100&13&-7.7&-2&0.82&73.1\\
AS&XR&294&09:22&48.5&35.5&10.8547&9&-7.5&19&0.60&53.4\\
AS&XR&294&09:30&46.8&37.3&10.8874&34&-7.9&-0&0.64&56.6\\
AS&XR&294&09:38&45.1&39.1&10.9100&9&-7.5&-2&0.79&70.8\\
AS&YL&294&09:22&48.5&35.5&10.8547&13&0.8&19&0.34&29.6\\
AS&YL&294&09:30&46.8&37.3&10.8874&17&0.4&0&0.47&41.3\\
AS&YL&294&09:38&45.1&39.1&10.9100&15&0.7&-2&0.51&45.2\\
AS&YR&294&09:22&48.5&35.5&10.8547&10&-5.9&19&0.46&40.7\\
AS&YR&294&09:30&46.8&37.3&10.8874&14&-6.3&0&0.50&44.2\\
AS&YR&294&09:38&45.1&39.1&10.9100&10&-5.9&-3&0.62&54.9\\
AX&XR&294&09:22&48.5&48.5&0.0028&27&-1.0&-8&0.14&12.6\\
AX&XR&294&09:30&46.8&46.8&0.0028&39&-0.9&-9&0.16&13.0\\
AX&XR&294&09:38&45.1&45.1&0.0028&32&-0.9&-11&0.15&12.9\\
AX&YR&294&09:22&48.5&48.5&0.0028&30&0.6&-7&0.14&10.9\\
AX&YR&294&09:30&46.8&46.8&0.0028&29&0.7&-9&0.14&10.8\\
\hline
\multicolumn{11}{c}{\jfiveone}\\
\hline
AS&XL&294&10:01&37.0&39.6&10.9218&30&-8.0&-12&0.10&8.5\\
AS&XR&294&10:01&37.0&39.6&10.9218&28&-8.0&-12&0.25&22.3\\
AS&XR&294&10:17&34.5&43.4&10.8891&8&-8.1&-0&0.27&22.4\\
AS&XR&294&10:22&33.5&44.8&10.8682&22&2.2&20&0.20&16.6\\
AS&YL&294&10:01&37.0&39.6&10.9218&10&0.3&-12&0.20&18.1\\
AS&YL&294&10:17&34.5&43.4&10.8891&23&0.2&11&0.25&21.3\\
AS&YL&294&10:22&33.5&44.8&10.8682&29&-6.6&2&0.17&14.2\\
AS&YR&294&10:01&37.0&39.6&10.9218&28&-6.3&-14&0.12&10.1\\
AS&YR&294&10:17&34.5&43.4&10.8891&6$^{**}$&-6.5&0&0.14&11.5\\
AS&YR&294&10:22&33.5&44.8&10.8682&10$^{**}$&3.8&81&0.11&9.7\\
\hline
\multicolumn{11}{c}{\jfivetwo}\\
\hline
AS&XR&294&10:37&53.0&18.0&10.3188&12$^{**}$&-4.7&38&0.12&10.1\\
AS&XR&294&10:45&51.4&19.2&10.4084&24&-4.9&8&0.20&12.1\\
AS&YL&294&10:37&53.0&18.0&10.3188&29&3.5&-4&0.12&10.3\\
AS&YL&294&10:45&51.4&19.2&10.4084&22&3.4&-4&0.16&14.1\\
AX&XR&294&10:37&53.0&52.9&0.0030&31&0.8&-1&0.31&26.9\\
AX&XR&294&10:45&51.4&51.4&0.0030&39&0.8&25&0.25&15.3\\
AX&YR&294&10:37&53.0&52.9&0.0030&31&2.3&1&0.31&27.0\\
AX&YR&294&10:45&51.4&51.4&0.0030&31&2.4&25&0.29&24.6\\
\enddata
\tablenotetext{$\dag$}{Baselines: AX (ALMA-APEX), AP (ALMA-IRAM30m), AS (ALMA-SMA)}
\tablenotetext{*}{Day of Year in 2018.}
\tablenotetext{**}{The S/N was insufficient to fit the coherence time.  The reported value is the segmentation time that achieves the greatest S/N for the scan.}
\end{deluxetable*}
%======================================

\subsection{1.3~mm (230 GHz) Comparison}
\begin{figure}
    \centering
    \includegraphics[width=\columnwidth]{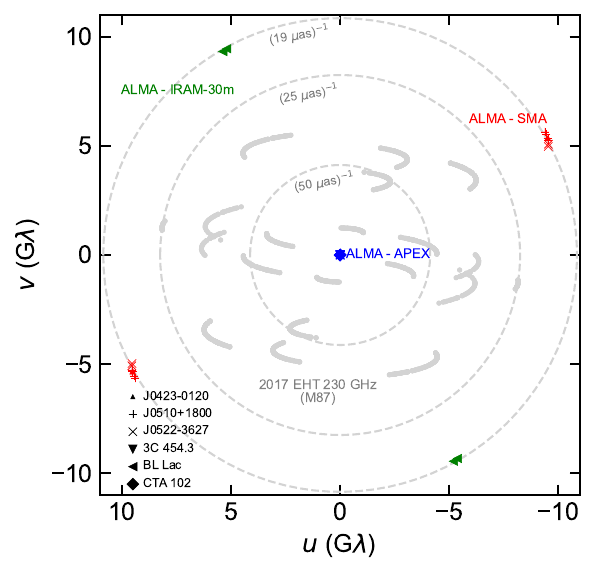}
    \caption{Detections on various targets at 345 GHz (see Table \ref{table:experiments345}). The u--v locations of 230 GHz detections on \m87 during the EHT April 2017 campaign are shown in gray including low-S/N scans at $\left(25 \mu \mathrm{as} \right)^{--1}$. 
    }
    \label{fig:uvplot}
\end{figure}

Presently, the EHT observes at 1.3~mm~\citep{paperii}.  Figure \ref{fig:uvplot} compares the Fourier components of the 870\um detections on various sources to the 1.3~mm coverage of the 2017 EHT array on \m87 \citep{paperiii}.  The 870\um detections on ALMA-IRAM30m and ALMA-SMA baselines have a higher nominal angular resolution (19\uas) than the highest-resolution \m87 detections (nominally 25 $\mu$as).

For a source-specific comparison of the 1.3~mm and 870\um bands, ALMA and IRAM30m observed \bllac at 1.3~mm during three scans at the end of the eastern subarray scheduling block of the October 2018 session.  Those data were searched using the same HOPS incoherent averaging method as was used for the 870\um observations and provide an independent application of the approach.   The 1.3~mm scans provide a check of the 870\um processing and a point of comparison for the 870\um detections.

\begin{figure}[h!]
\centering
  \includegraphics[width=1\columnwidth]{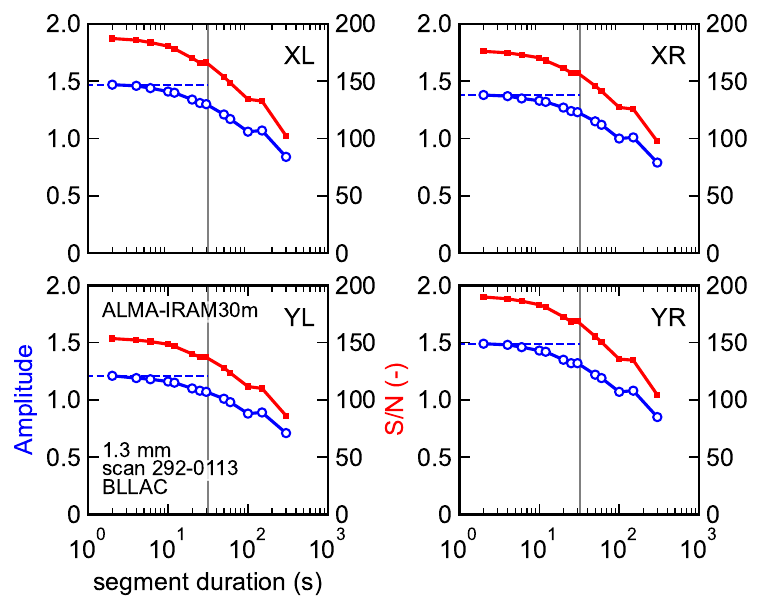}
  \caption{1.3~mm amplitude (open blue circles, left axes) and $S/N$ (closed red squares, right axes) versus the duration of coherently integrated segments, which are incoherently averaged.  Each subplot shows a different polarization on the baseline between ALMA and IRAM30m for a single scan on \bllac on October 19, 01:13 UTC.  Other \bllac detections listed in Table~\ref{table:experiments230} have similar dependence on segment duration.  The noise-debiased amplitude and coherence time were derived using HOPS and are indicated by the horizontal blue dashed line and the vertical solid black line, respectively.  These data were calibrated in the same manner as the 870\um detections.}
  \label{fig:coherencetime230}
\end{figure}

The amplitude and $S/N$ values for one of the 1.3~mm scans are plotted in Fig.~\ref{fig:coherencetime230} versus the duration of incoherently-averaged segments.  The $S/N$ values are approximately 10-fold greater at 1.3~mm than at 870\um (see Figure~\ref{fig:coherencetime}), which likely results from a combination of factors that boost sensitivity at the longer wavelength: lower opacity, lower receiver noise, greater aperture efficiency, a wider beam, greater coherence, and greater source flux density.  The coherence time determined using HOPS was comparable for the three scans to what was found at 870\um: on the order of 6 to 30 seconds. As with the 870\um measurements, the $S/N$ values asymptote as the segment duration decreases below the coherence time.  The consistency of the $S/N$ trends in the 870\um and 1.3~mm scans suggests that the behavior is a real feature of the data and not an artifact of the analysis.

Comparison of the 1.3~mm and 870\um wavelengths observing \bllac also shows that the latter is a much more difficult regime in which to operate. The atmospheric conditions at the IRAM30m site (see Fig.~\ref{fig:experiment}; $\tau_{345}\sim0.8$) were not ideal for 870\um observing during the test.  At 1.3~mm, strong detections were obtained on all polarizations for each of the three attempted scans.  At 870\um, detections were made on just two of four polarizations for a single ALMA-IRAM30m scan, and none were made on other \bllac baselines.  The 10-fold greater $S/N$ values at 1.3~mm are consistent with the system equivalent flux density (SEFD).  The SEFD on \bllac scans at ALMA were approximately 150~Jy at 1.3~mm versus 580~Jy at 870\um (factor of 3.9 change).  At IRAM30m, SEFDs during the \bllac scans were 3800~Jy at 1.3~mm versus 10$^5$~Jy at 870\um (factor of approximately 25 change).  The $S/N$ is inversely proportional to the root product of the SEFDs, or $\sqrt{3.9 \times 25} \approx 10$, which explains the behavior across observing wavelengths.  The significantly greater noise at 870\um as well as the other losses associated with narrower beam width or coherence is the likely reason for non-detections to some stations and on certain scans.

% =============== FIGURE 05 ==========
% =========================

\begin{figure}[h!]
\centering
  \includegraphics[width=1\columnwidth]{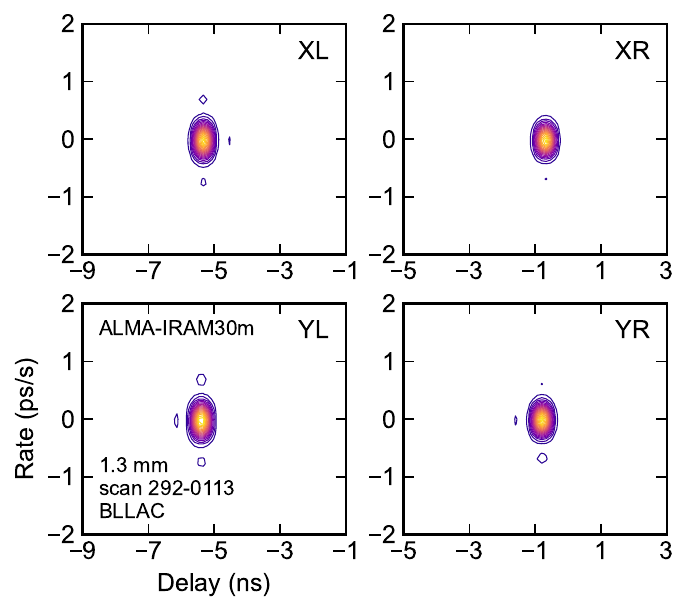}
  \caption{1.3~mm contours of incoherently-averaged fringe power in 5\% increments versus delay and rate for the baseline between ALMA and IRAM30m. This example is for a single scan on \bllac taken on October 19, 01:13 UTC.  Other detections reported in Table~\ref{table:experiments230} also exhibit clear peaks versus in delay-delay rate search space.}
  \label{fig:delayrate230}
\end{figure}

Fringe power contours at 1.3~mm are plotted as a function of multi-band delay and rate in Fig.~\ref{fig:delayrate230}, exhibiting obvious peaks.  The delays for each of the four polarization cross products is consistent across scans, and the 1.3~mm fringes are summarized in Table~\ref{table:experiments230}.  All four polarization cross-hands are detected in each of the three 1.3~mm scans.  The 6.4~G$\lambda$ spatial frequencies are 50\% smaller than the 870\um scans on the AP baseline, which corresponds to the frequency scaling between the two bands.  The 1.3~mm zero-baseline flux density of \bllac deduced from the ALMA local interferometry~\citep{crew2023} was 1.2~Jy.

%=========FIGURE 6===========
%============================

%=========FIGURE 7===========
%============================

%=========TABLE 2===========
\begin{deluxetable}{ccccccc}
\tablecaption{1.3~mm detections on the ALMA -- IRAM-30~m baseline toward \bllac for indicated polarizations. Scans listed top to bottom on October 19 begin at 01:03, 01:09, and 01:13 UTC. \label{table:experiments230}}
\tabletypesize{\footnotesize}
\tablehead{
\colhead{\textbf{Elevation}} &
\colhead{\textbf{Baseline}} &
\colhead{\textbf{$\tau_c$}} &
 \colhead{\textbf{Delay}} &
 \colhead{\textbf{Rate}} &
\colhead{\textbf{Amp.}} &
\colhead{\textbf{S/N}} \\
\colhead{\textbf{(ALMA/IRAM30m)}} &
 \colhead{\textbf{Length}} &
\colhead{} &
 \colhead{} &
 \colhead{} &
\colhead{} &
\colhead{} \\
\colhead{(\degr)} &
 \colhead{(G$\lambda$)} &
\colhead{(s)} &
 \colhead{(ns)} &
 \colhead{(fs~s$^{-1}$)} &
\colhead{($\times10^{-4}$)} &
\colhead{} 
}
\startdata
\multicolumn{7}{c}{\textit{XL}}\\
\hline
24.5 / 38.3&6.4327&5&-5.3&-98&1.66&134.0\\
24.4 / 37.3&6.4422&7&-5.3&-66&1.49&120.0\\
24.3 / 36.6&6.4476&32&-5.3&-14&1.47&187.3\\
\hline
\multicolumn{7}{c}{\textit{YR}}\\
\hline
24.5 / 38.3&6.4327&6&-0.8&-99&1.77&143.0\\
24.4 / 37.3&6.4422&7&-0.8&-66&1.52&122.4\\
24.3 / 36.6&6.4476&32&-0.8&-14&1.49&189.8\\
\hline
\multicolumn{7}{c}{\textit{XR}}\\
\hline
24.5 / 38.3&6.4327&6&-0.7&-98&1.56&125.4\\
24.4 / 37.3&6.4422&7&-0.7&-66&1.37&110.4\\
24.3 / 36.6&6.4476&32&-0.7&-13&1.38&176.1\\
\hline
\multicolumn{7}{c}{\textit{YL}}\\
\hline
24.5 / 38.3&6.4327&6&-5.4&-98&1.42&114.4\\
24.4 / 37.3&6.4422&7&-5.4&-66&1.24&100.1\\
24.3 / 36.6&6.4476&32&-5.4&-14&1.21&153.5\\
\enddata
\end{deluxetable}

\subsection{Coherence and Allan Deviation}  

It is convenient to characterize the phase noise of an interferometer by its Allan deviation, which is a measure of fractional stability for an oscillator, time standard or any time variable process.   When computing the Allan deviation of observed VLBI interferometer phase one normalizes by the frequency of observation to produce a dimensionless quantity. The relationships of Allan deviation to the statistical variance, coherence, and phase power spectrum can be found in \cite{TMS2017}. Examples of the Allan deviation of VLBI systems referenced to hydrogen maser time standards and operating at 1.3 cm and 3 mm wavelength are can be found in \cite{rogers1981} and \cite{rogers1984} respectively, and show that at short wavelengths decoherence is a potential concern.  Alternatives to hydrogen masers for short-wavelength VLBI work have been explored (e.g., ~\citet{Doeleman_2011}).  In this section we compare the observed Allan deviation of the VLBI interferometric phase to limiting factors including the stability of time and frequency standards used in the experiment as well as instabilities due to atmospheric turbulence.

%=========FIGURE 8===========
\begin{figure}[ht]
\centering
  \includegraphics[width=1.\columnwidth]{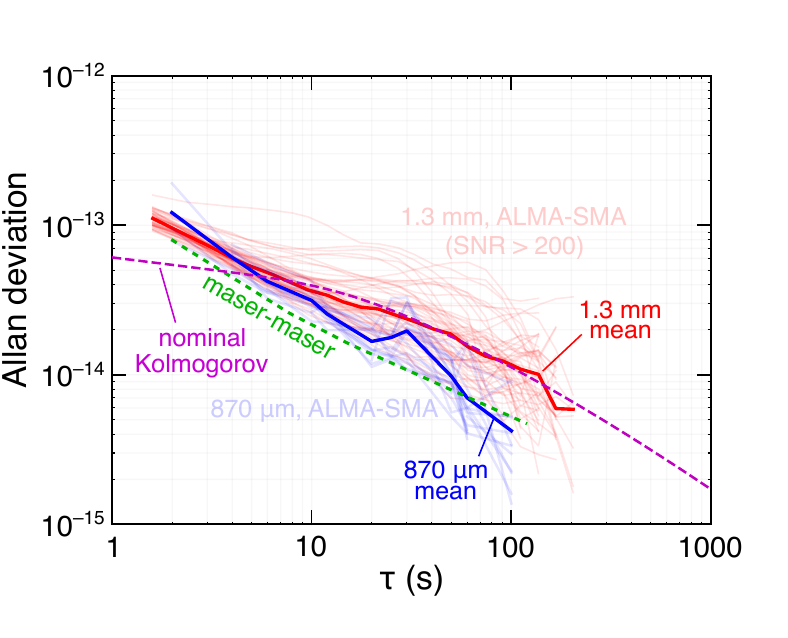}
  \caption{Allan deviation for 870\um (345~GHz) scans observed on the ALMA-SMA baseline (blue traces). For comparison, red traces show the Allan deviation for high-$S/N$ scans (nominally 5 minutes long) during the 1.3~mm (230~GHz) 2017 EHT campaign~\citep{paperii}.  Weather variability during the 2017 campaign is responsible for the spread in those scans.    The means of the individual Allan deviation traces are shown in bold for the two frequencies.  The 870\um and 1.3~mm mean traces approach the nominal Allan deviation for a pair of T4 Science brand iMaser 3000 model masers~\citep{TMS2017} at short timescales.  At intermediate timescales, atmospheric turbulence can become important.  The Allan deviation associated with Kolmogorov turbulence is plotted for a set of nominal parameters~\citep{treuhaft1987}.}
  \label{fig:adev}
\end{figure}
%============================

Figure~\ref{fig:adev} shows the Allan deviation for 870\um scans on the ALMA-SMA baseline.  Over most integration times, the 870\um Allan deviation is comparable to but greater than the maser-maser reference.  The 870\um traces exhibit relatively small scan-to-scan variation during the course of the brief fringe test when conditions were relatively stable.  For comparison, Fig.~\ref{fig:adev} also shows the Allan deviations for a large number of high-$S/N$ 1.3~mm scans from the 2017 EHT campaign~\citep{paperiii}.  At times less than about 5~s, the red 1.3~mm traces all approach the limit set by the maser references.  At times longer than 5~s, the red traces are noticeably scattered.  The scatter exists because of the variability of atmospheric conditions during the course of an observing campaign.  

The tropospheric delay is essentially independent of wavelength for wavelengths longer than about 600\um as described by the Smith-Weintraub  equation (see \cite{TMS2017}, chapter 13).  Thus the Allan deviation is expected to be independent of wavelength for our observations.   When the atmospheric conditions are stable the 1.3~mm Allan deviation for individual scans approaches the maser-maser limit across all integration times.  The mean of the 1.3~mm scans is within a factor of approximately two of the mean of the 870\um traces.  The 870\um mean Allan deviation on the plot happens to be lower than the 1.3~mm mean for most integration times.  However we do not consider this difference to be significant give the extremely small 870\um data set. Further, the observations in 2017 April and 2018 October observations were of course made in differing weather conditions.

To assess the impact of atmospheric turbulence at longer times, the Allan deviation associated with atmospheric Kolmogorov turbulence is plotted for a set of nominal conditions following the approach outlined by~\cite{treuhaft1987}: 10~m~$\mathrm{s}^{-1}$ wind speed, 2~km troposphere scale height, 1.99$\times 10^{-7}$~m$^{-1/3}$ Kolmogorov coefficient, and independent distant sites.  The nominal Kolmogorov trace exceeds the maser-maser Allan deviation at longer times where we expect atmospheric effects to dominate. Beyond  10~s, the nominal Kolmogorov trace matches the shape of the 1.3~mm mean.  Although the 870\um mean falls somewhere between the maser-maser and nominal Kolmogorov limits, the atmospheric contribution may become more apparent the future with scans spanning more variable weather conditions.

\subsection{Phasing Efficiency} \label{sec:phasing}

An important figure of merit when used to monitor the performance of phased array beamformers is phasing efficiency. This is a measure of how effectively outputs of the dishes in the local array are coherently summed to synthesize a single IF output from the array's aggregated collecting area. For each array site periodic estimates of phasing efficiency over time are stored with other essential metadata for use in calibration.

The ALMA and SMA phased arrays experienced lower and more variable phasing efficiency during the 870\um test than is typical for 1.3\,mm observing in similar conditions. At 870\um atmospheric opacity is between 3 and 3.5 times that for 1.3\,mm given the same precipitable  water vapor (PWV).  Further source fluxes decline with increasing frequency or shorter wavelength. Both of these factors result in lower local array fringe signal-to-noise-ratio (SNR). There is thus greater error in the fits of the antenna phase corrections. Tuning within the band avoids the deep absorption lines due to atmospheric water resonances at 325 and 385\,GHz which would reduce the SNR still further. Also, the atmospheric phase fluctuations tracked by the adaptive phased array system have a greater amplitude for observations in the higher frequency band. \cite{crew2023} note that that moist, windy conditions tend to diminish phasing efficiency, and the  winds were quite high at ALMA during the test.  At dry less windy times ALMA obtained higher phasing efficiencies approaching 100\%.  While NOEMA participated in this test with a single dish, not as a phased array, all of these factors are expected to apply as well to NOEMA which is now equipped with a phased array back end capable of beamforming in both the 1.3\,mm and 870\um bands.

Water vapor radiometer (WVR) based  phasing corrections were not in use during the 2018 test. Independent testing at ALMA show that fast WVR corrections are  effective at  improving the efficiency when phase fluctuations are primarily due to water vapor.  Phasing control loop algorithms are constantly being improved and in future will be better tuned to the 870\um waveband. These improvements will expand the opportunities for 870\um observing in a wider range of weather conditions and on weaker sources. Despite these challenges VLBI detections at 870\um can be readily achieved even when phasing efficiencies are relatively low and in non-ideal weather conditions.

\vspace{0.5cm}
\section{Future Directions}
% =============== FIGURE 09 ==========
\begin{figure*}
\centering
  \includegraphics[width=\textwidth]{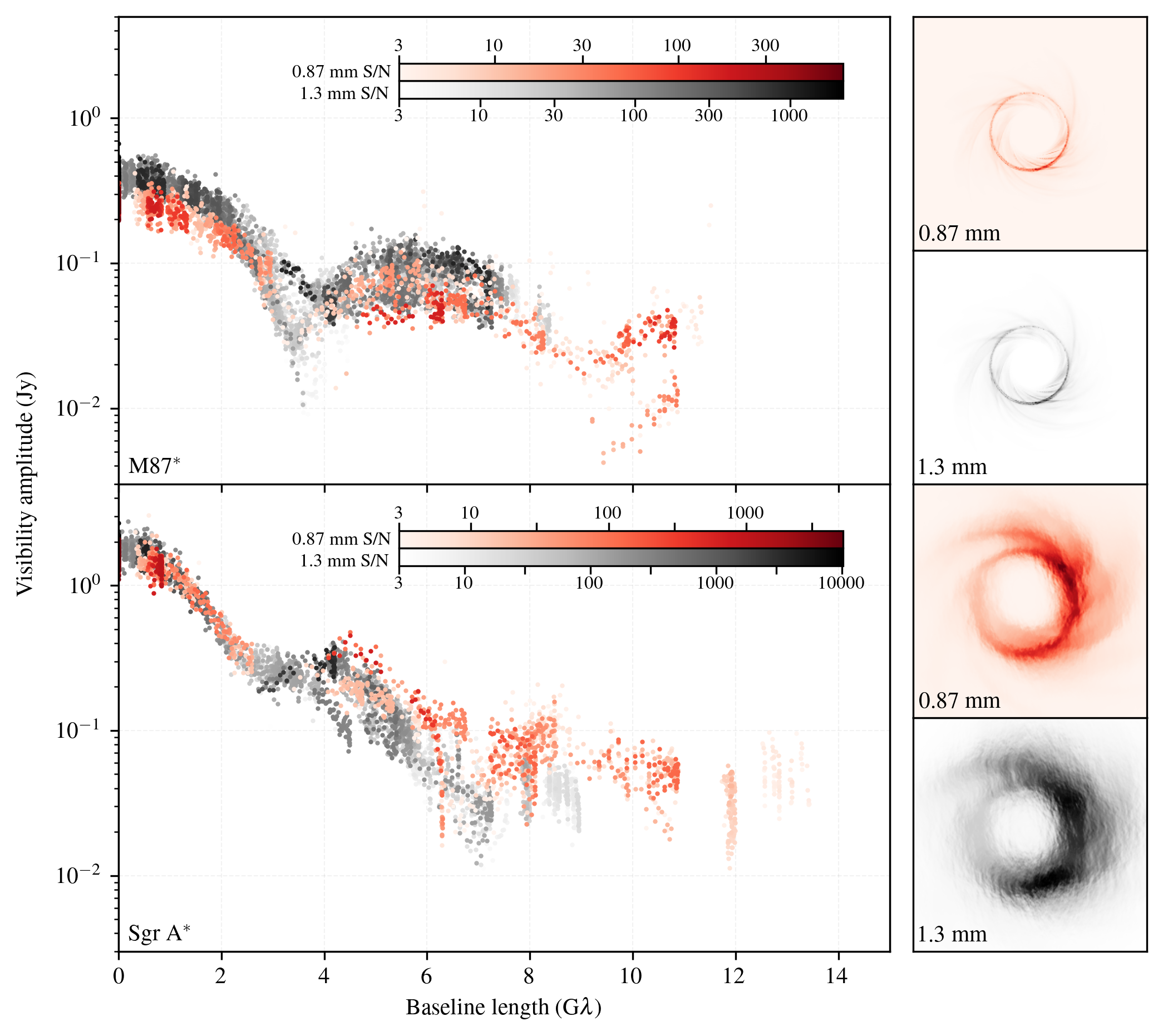}
  \caption{\textit{Left}: Visibility amplitudes for simulated observations of \m87 (top) and \sgra (bottom) at observing wavelengths of 1.3\,mm (gray) and 0.87\,mm (red).  The synthetic data have been generated using the \texttt{ngehtsim} package assuming array specifications appropriate for the Phase 2 next-generation EHT array from \citet{ngeht_refarray}, including simultaneous dual-band observations, the use of the frequency phase transfer calibration technique, and 16\,GHz of bandwidth at both frequencies.  Data points are colored by their S/N on an integration time of 5 minutes, and data points with $\text{S/N} < 3$ have been flagged.  \textit{Right}: Images produced from GRMHD simulations of the \m87 \citep[top two panels;][]{M87PaperV} and \sgra \citep[bottom two panels;][]{SgrAPaperV} accretion flows, used to generate the synthetic data shown in the left panels.  Both simulations have been ray-traced at observing wavelengths of 1.3\,mm (gray) and 0.87\,mm (red), and the frequency-dependent effects of interstellar scattering have been applied to the \sgra images \citep{johnson2016,johnson2018}.}
  \label{fig:model}
\end{figure*}
% =========================
Achieving 870\um VLBI fringes has strong implications for science directions that future global arrays operating at this wavelength can explore.  As angular resolution scales with wavelength, we anticipate improving resolution from $\sim23\;\mu$as to $\sim15\;\mu$as on the longest EHT baselines (Figure 5).  Plasma propagation processes typically scale as wavelength squared, so at 870\um scatter broadening of \sgra reduces to $\sim5\;\mu$as, further sharpening resolution and increasing signal-to-noise on the longest VLBI baselines.  Similarly, Faraday Rotation measured across the bandpass of EHT receivers at 870\um can be used to improve estimates of accretion plasma densities and magnetic field geometries close to EHT targets.  For both Sgr A* and M87* the images at 870\um and 1.3~mm are determined predominantly by the achromatic gravitational lensing, and hence should exhibit similar characteristics, implying that the aggregate Fourier coverage of VLBI observations at different frequencies can be used to improve modeling of the gravitationally lensed emission, and imaging fidelity generally \citep{Chael_MFS_2023}.  Figure~\ref{fig:model} shows Fourier amplitudes as a function of radius for GRMHD\footnote{General Relativistic Magnetohydrodynamic} models of \m87 and \sgra.  Inclusion of 345~GHz observations adds coverage in the visibility plane regions not sampled at 230~GHz, and it extends baseline lengths for higher angular resolution as well as enhanced overall sampling of Fourier spatial frequencies to allow dynamical reconstructions of accretion and jet launch close to the event horizon.

There are several developments that will increase the sensitivity and flexibility of 870$\nu$m VLBI in the near future.  Next-generation VLBI backends \citep{ngeht_refarray} will allow an increase in data capture rates from 64 to 128~Gb/s (per observing frequency band), lowering detection thresholds by $\sqrt{2}$.  Additional use of the Frequency Phase Transfer technique \citep[FPT;][]{ngeht_fpt} through simultaneous observations at 86, 230 and 345GHz will extend coherent integration times at higher frequencies, further increasing sensitivity. In optimal cases this increase will be the square root of the ratio of coherence times at 86GHz and 345GHz ($\sqrt{\tau_c(86)/\tau_c(345)}$).  And the participation of more telescopes at high altitude sites will make the EHT array more robust against adverse weather conditions, increasing the opportunities for staging 870\um VLBI observations \citep{raymond2021,ngeht_refarray}.  Anticipated upgrades to the ALMA array will be exceptionally useful to advance 870\um VLBI, and are planned on a similar timeline ($\sim$2030) as the ngEHT upgrade \citep{ALMA_2030_WSU}.  In particular, the projected doubling of continuum bandwidth of ALMA will match the ngEHT specifications, and a sub-array capability at ALMA will enable simultaneous multi-band observations that benefit from FPT as noted above.  In sum, the prospects for routine 870\um VLBI in the near future are excellent.

\section{Conclusions}
VLBI fringe detections on baselines between ALMA-APEX, ALMA-IRAM30m, and ALMA-SMA have been achieved at 870\um for multiple AGN sources.  Signal-to-noise ratios were between approximately 10 and 70.  Despite marginal weather conditions across the array, detections to multiple stations and sources were obtained.  This work demonstrates that the EHT instrumentation is viable at 870\um (345~GHz) and will provide a critical advance in array capability.  EHT-wide observations at 870\um would yield a fringe spacing of about 15\uas, and with a full-track of coverage, would significantly enhance the fine details of the EHT images of AGN and horizon-scale targets~\citep{Doeleman_2019,ngeht_refarray,ngeht_ksg}.

\acknowledgments
 This work was supported by the National Science Foundation (grants AST-1935980, AST-1743747, AST-1440254), an ALMA Cycle 5 North America Development Award, the Gordon and Betty Moore Foundation (awards GBMF3561, GBMF5278, and GBMF10423), and a generous gift from the Deepak Raghavan Family Foundation.  Work on this project was conducted in part at the Black Hole Initiative at Harvard University (funded through grants 60477, 61479 and 62286 from the John Templeton Foundation; and grant GBMF8273 from the Gordon and Betty Moore Foundation).  This work is partly based on observations carried out with the IRAM 30-m telescope and the NOEMA Interferometer. IRAM is supported by INSU/CNRS (France), MPG (Germany) and IGN (Spain). The IRAM NOEMA phasing project was supported by the European Research Council (ERC) Synergy Grant "BlackHoleCam: Imaging the Event Horizon of Black Holes" (grant 610058). The Submillimeter Array is a joint project between the Smithsonian Astrophysical Observatory and the Academia Sinica Institute of Astronomy and Astrophysics and is funded by the Smithsonian Institution and the Academia Sinica. The SMA gratefully acknowledges the efforts of its staff for supporting these observations, including those of the operator on duty, R. Howie.  This publication is based on data acquired with the Atacama Pathfinder Experiment (APEX). APEX is a collaboration between the Max-Planck-Institut fur Radioastronomie, the European Southern Observatory, and the Onsala Space Observatory.This work was an activity external to JPL, and effort by A.R. was not in their capacity as an employee of the Jet Propulsion Laboratory, California Institute of Technology.  The Greenland Telescope (GLT) is supported by the the Ministry
of Science and Technology (MOST) of Taiwan (103-2119-M-001-010-MY2,
105-2119-M-001-042, 106-2119-M-001-013, 107-2119-M-001-041,
108-2112-M-001-048, 109-2124-M-001-005, 110-2124-M-001-007) and the
National Science and Technology Council (NSTC) of Taiwan
(111-2124-M-001-005, 112-2124-M-001-014).  

%\begin{acknowledgments}
The Event Horizon Telescope Collaboration additionally thanks the following
organizations and programs: the Academia Sinica; the Academy
of Finland (projects 274477, 284495, 312496, 315721); the Agencia Nacional de Investigaci\'{o}n 
y Desarrollo (ANID), Chile via NCN$19\_058$ (TITANs), Fondecyt 1221421 and BASAL FB210003; the Alexander
von Humboldt Stiftung; an Alfred P. Sloan Research Fellowship;
Allegro, the European ALMA Regional Centre node in the Netherlands, the NL astronomy
research network NOVA and the astronomy institutes of the University of Amsterdam, Leiden University, and Radboud University;
the ALMA North America Development Fund; the Astrophysics and High Energy Physics programme by MCIN (with funding from European Union NextGenerationEU, PRTR-C17I1); 
%the Black Hole Initiative, which is funded by grants from the John Templeton Foundation (60477, 61497, 62286) and the Gordon and Betty Moore Foundation (Grant GBMF-8273) - although the opinions expressed in this work are those of the author and do not necessarily reflect the views of these Foundations; %the Black Hole Initiative, which is funded by grants from the John Templeton Foundation and the Gordon and Betty Moore Foundation (although the opinions expressed in this work are those of the author(s) and do not necessarily reflect the views of these Foundations); 
the Brinson Foundation; ``la Caixa'' Foundation (ID 100010434) through fellowship codes LCF/BQ/DI22/11940027 and LCF/BQ/DI22/11940030; 
Chandra DD7-18089X and TM6-17006X; the China Scholarship
Council; the China Postdoctoral Science Foundation fellowships (2020M671266, 2022M712084); Consejo Nacional de Humanidades, Ciencia y Tecnología (CONAHCYT, Mexico, projects U0004-246083, U0004-259839, F0003-272050, M0037-279006, F0003-281692, 104497, 275201, 263356); the Colfuturo Scholarship; 
the Consejer\'{i}a de Econom\'{i}a, Conocimiento, 
Empresas y Universidad 
of the Junta de Andaluc\'{i}a (grant P18-FR-1769), the Consejo Superior de Investigaciones 
Cient\'{i}ficas (grant 2019AEP112);
the Delaney Family via the Delaney Family John A.
Wheeler Chair at Perimeter Institute; Dirección General de Asuntos del Personal Académico-Universidad Nacional Autónoma de México (DGAPA-UNAM, projects IN112820 and IN108324); the Dutch Research Council (NWO) for the VICI award (grant 639.043.513), the grant OCENW.KLEIN.113, and the Dutch Black Hole Consortium (with project No. NWA 1292.19.202) of the research programme the National Science Agenda; the Dutch National Supercomputers, Cartesius and Snellius  
(NWO grant 2021.013); 
the EACOA Fellowship awarded by the East Asia Core
Observatories Association, which consists of the Academia Sinica Institute of Astronomy and
Astrophysics, the National Astronomical Observatory of Japan, Center for Astronomical Mega-Science,
Chinese Academy of Sciences, and the Korea Astronomy and Space Science Institute; 
%the European Research Council (ERC) Synergy Grant ``BlackHoleCam: Imaging the Event Horizon of Black Holes" (grant 610058); 
the European Union Horizon 2020
research and innovation programme under grant agreements
RadioNet (No. 730562), 
M2FINDERS (No. 101018682) and FunFiCO (No. 777740); the European Research Council for advanced grant `JETSET: Launching, propagation and 
emission of relativistic jets from binary mergers and across mass scales' (grant No. 884631); the European Horizon Europe staff exchange (SE) programme HORIZON-MSCA-2021-SE-01 grant NewFunFiCO (No. 10108625); the Horizon ERC Grants 2021 programme under grant agreement No. 101040021; the FAPESP (Funda\c{c}\~ao de Amparo \'a Pesquisa do Estado de S\~ao Paulo) under grant 2021/01183-8; the Fondo CAS-ANID folio CAS220010; 
the Generalitat
Valenciana (grants APOSTD/2018/177 and  ASFAE/2022/018) and
GenT Program (project CIDEGENT/2018/021); 
%the Gordon and Betty Moore Foundation (GBMF-3561, GBMF-5278, GBMF-10423);   
the Institute for Advanced Study; the Istituto Nazionale di Fisica
Nucleare (INFN) sezione di Napoli, iniziative specifiche
TEONGRAV; 
the International Max Planck Research
School for Astronomy and Astrophysics at the
Universities of Bonn and Cologne; 
DFG research grant ``Jet physics on horizon scales and beyond'' (grant No. 443220636);
Joint Columbia/Flatiron Postdoctoral Fellowship (research at the Flatiron Institute is supported by the Simons Foundation); 
the Japan Ministry of Education, Culture, Sports, Science and Technology (MEXT; grant JPMXP1020200109); %the Japanese Government (Monbukagakusho:MEXT) Scholarship; 
the Japan Society for the Promotion of Science (JSPS) Grant-in-Aid for JSPS
Research Fellowship (JP17J08829); the Joint Institute for Computational Fundamental Science, Japan; the Key Research
Program of Frontier Sciences, Chinese Academy of
Sciences (CAS, grants QYZDJ-SSW-SLH057, QYZDJSSW-SYS008, ZDBS-LY-SLH011); 
the Leverhulme Trust Early Career Research
Fellowship; the Max-Planck-Gesellschaft (MPG);
the Max Planck Partner Group of the MPG and the
CAS; the MEXT/JSPS KAKENHI (grants 18KK0090, JP21H01137,
JP18H03721, JP18K13594, 18K03709, JP19K14761, 18H01245, 25120007, 19H01943, 21H01137, 21H04488, 22H00157, 23K03453); the MICINN Research Projects PID2019-108995GB-C22, PID2022-140888NB-C22; the MIT International Science
and Technology Initiatives (MISTI) Funds; 
the Ministry of Science and Technology (MOST) of Taiwan (105-2112-M-001-025-MY3, 106-2112-M-001-011, 106-2119-M-001-027, 106-2923-M-001-005, 107-2119-M-001-017, 107-2119-M-001-020, 107-2119-M-110-005, 107-2923-M-001-009, 108-2112-M-001-051, 108-2923-M-001-002, 109-2112-M-001-025, 109-2923-M-001-001, %110-2112-M-003-007-MY2, 
110-2112-M-001-033, 110-2923-M-001-001, and 112-2112-M-003-010-MY3);
the Ministry of Education (MoE) of Taiwan Yushan Young Scholar Program;
the Physics Division, National Center for Theoretical Sciences of Taiwan;
the National Aeronautics and
Space Administration (NASA, Fermi Guest Investigator
grant %80NSSC20K1567
80NSSC23K1508, NASA Astrophysics Theory Program grant 80NSSC20K0527, NASA NuSTAR award 
80NSSC20K0645); 
NASA Hubble Fellowship 
grants HST-HF2-51431.001-A, HST-HF2-51482.001-A awarded 
by the Space Telescope Science Institute, which is operated by the Association of Universities for 
Research in Astronomy, Inc., for NASA, under contract NAS5-26555; 
the National Institute of Natural Sciences (NINS) of Japan; the National
Key Research and Development Program of China
(grant 2016YFA0400704, 2017YFA0402703, 2016YFA0400702); the National Science and Technology Council (NSTC, grants NSTC 111-2112-M-001 -041, NSTC 111-2124-M-001-005, NSTC 112-2124-M-001-014); the US National
Science Foundation (NSF, grants AST-0096454,
AST-0352953, AST-0521233, AST-0705062, AST-0905844, AST-0922984, AST-1126433, OIA-1126433, AST-1140030,
DGE-1144085, AST-1207704, AST-1207730, AST-1207752, MRI-1228509, OPP-1248097, AST-1310896, AST-1555365, AST-1614868, AST-1615796, AST-1715061, AST-1716327,  AST-1726637, %AST-1716536, 
OISE-1743747, AST-1816420, AST-1952099, AST-2034306,  AST-2205908, AST-2307887); 
NSF Astronomy and Astrophysics Postdoctoral Fellowship (AST-1903847); 
the Natural Science Foundation of China (grants 11650110427, 10625314, 11721303, 11725312, 11873028, 11933007, 11991052, 11991053, 12192220, 12192223, 12273022, 12325302, 12303021); 
the Natural Sciences and Engineering Research Council of
Canada (NSERC, including a Discovery Grant and
the NSERC Alexander Graham Bell Canada Graduate
Scholarships-Doctoral Program); %the National Youth Thousand Talents Program of China; 
the National Research
Foundation of Korea (the Global PhD Fellowship
Grant: grants NRF-2015H1A2A1033752, the Korea Research Fellowship Program:
NRF-2015H1D3A1066561, Brain Pool Program: 2019H1D3A1A01102564, 
Basic Research Support Grant 2019R1F1A1059721, 2021R1A6A3A01086420, 2022R1C1C1005255, 2022R1F1A1075115); 
Netherlands Research School for Astronomy (NOVA) Virtual Institute of Accretion (VIA) postdoctoral fellowships; NOIRLab, which is managed by the Association of Universities for Research in Astronomy (AURA) under a cooperative agreement with the National Science Foundation; 
Onsala Space Observatory (OSO) national infrastructure, for the provisioning
of its facilities/observational support (OSO receives
funding through the Swedish Research Council under
grant 2017-00648);  the Perimeter Institute for Theoretical
Physics (research at Perimeter Institute is supported
by the Government of Canada through the Department
of Innovation, Science and Economic Development
and by the Province of Ontario through the
Ministry of Research, Innovation and Science); the Portuguese Foundation for Science and Technology (FCT) grants (Individual CEEC program - 5th edition, \url{https://doi.org/10.54499/UIDB/04106/2020}, \url{https://doi.org/10.54499/UIDP/04106/2020}, PTDC/FIS-AST/3041/2020, CERN/FIS-PAR/0024/2021, 2022.04560.PTDC); the Princeton Gravity Initiative; the Spanish Ministerio de Ciencia e Innovaci\'{o}n (grants PGC2018-098915-B-C21, AYA2016-80889-P,
PID2019-108995GB-C21, PID2020-117404GB-C21); 
the University of Pretoria for financial aid in the provision of the new 
Cluster Server nodes and SuperMicro (USA) for a SEEDING GRANT approved toward these 
nodes in 2020; the Shanghai Municipality orientation program of basic research for international scientists (grant no. 22JC1410600); 
the Shanghai Pilot Program for Basic Research, Chinese Academy of Science, 
Shanghai Branch (JCYJ-SHFY-2021-013);
the State Agency for Research of the Spanish MCIU through
the ``Center of Excellence Severo Ochoa'' award for
the Instituto de Astrof\'{i}sica de Andaluc\'{i}a (SEV-2017-
0709); the Spanish Ministry for Science and Innovation grant CEX2021-001131-S funded by MCIN/AEI/10.13039/501100011033; the Spinoza Prize SPI 78-409; the South African Research Chairs Initiative, through the 
South African Radio Astronomy Observatory (SARAO, grant ID 77948),  which is a facility of the National 
Research Foundation (NRF), an agency of the Department of Science and Innovation (DSI) of South Africa; the Swedish Research Council (VR); the Taplin Fellowship; the Toray Science Foundation; the UK Science and Technology Facilities Council (grant no. ST/X508329/1); the US Department of Energy (USDOE) through the Los Alamos National
Laboratory (operated by Triad National Security,
LLC, for the National Nuclear Security Administration
of the USDOE, contract 89233218CNA000001); and the YCAA Prize Postdoctoral Fellowship.

We thank
the staff at the participating observatories, correlation
centers, and institutions for their enthusiastic support.  This work made use of the following ALMA data: ADS/JAO.ALMA\#2011.0.00010.E for the VLBI sessions and ADS/JAO.ALMA\#2011.0.00012.E for the Band 7 test data as well as ADS/JAO.ALMA\#2016.1.01154.V.  ALMA is a partnership of ESO (representing its member states), NSF (USA) and NINS (Japan), together with NRC (Canada), NSTC and ASIAA (Taiwan), and KASI (Republic of Korea), in cooperation with the Republic of Chile. The Joint ALMA Observatory is operated by ESO, AUI/NRAO and NAOJ.  See https://almascience.eso.org/almadata/ec/eht-2018 for more detail and access to the data.
%This paper makes use of the following ALMA data: ADS/JAO.ALMA\#2016.1.01154.V. 
%ALMA is a partnership of the European Southern Observatory (ESO; Europe, representing its member states), NSF, and National Institutes of Natural Sciences of Japan, together with National Research Council (Canada), Ministry of Science and Technology (MOST; Taiwan), Academia Sinica Institute of Astronomy and Astrophysics (ASIAA; Taiwan), and Korea Astronomy and Space Science Institute (KASI; Republic of Korea), in cooperation with the Republic of Chile. The Joint ALMA Observatory is operated by ESO, Associated Universities, Inc. (AUI)/NRAO, and the National Astronomical Observatory of Japan (NAOJ). 
The NRAO
is a facility of the NSF operated under cooperative agreement
by AUI.
This research used resources of the Oak Ridge Leadership Computing Facility at the Oak Ridge National
Laboratory, which is supported by the Office of Science of the U.S. Department of Energy under contract
No. DE-AC05-00OR22725; the ASTROVIVES FEDER infrastructure, with project code IDIFEDER-2021-086; the computing cluster of Shanghai VLBI correlator supported by the Special Fund 
for Astronomy from the Ministry of Finance in China;  
We also thank the Center for Computational Astrophysics, National Astronomical Observatory of Japan. This work was supported by FAPESP (Fundacao de Amparo a Pesquisa do Estado de Sao Paulo) under grant 2021/01183-8.

The EHTC has
received generous donations of FPGA chips from Xilinx
Inc., under the Xilinx University Program. The EHTC
has benefited from technology shared under open-source
license by the Collaboration for Astronomy Signal Processing
and Electronics Research (CASPER). The EHT
project is grateful to T4Science and Microsemi for their
assistance with hydrogen masers. This research has
made use of NASA's Astrophysics Data System. We
gratefully acknowledge the support provided by the extended
staff of the ALMA, from the inception of
the ALMA Phasing Project through the observational
campaigns of 2017 and 2018. We would like to thank
A. Deller and W. Brisken for EHT-specific support with
the use of DiFX. 
%We thank Martin Shepherd for the addition of extra features in the Difmap software that were used for the CLEAN imaging results presented in this paper.
We acknowledge the significance that Maunakea, where the SMA EHT station is located, has for the indigenous Hawaiian people.

%\end{acknowledgments}

\bibliographystyle{aasjournal}

\bibliography{references}

\begin{thebibliography}{}
\expandafter\ifx\csname natexlab\endcsname\relax\def\natexlab#1{#1}\fi
\providecommand{\url}[1]{\href{#1}{#1}}
\providecommand{\dodoi}[1]{doi:~\href{http://doi.org/#1}{\nolinkurl{#1}}}
\providecommand{\doeprint}[1]{\href{http://ascl.net/#1}{\nolinkurl{http://ascl.net/#1}}}
\providecommand{\doarXiv}[1]{\href{https://arxiv.org/abs/#1}{\nolinkurl{https://arxiv.org/abs/#1}}}

\bibitem[{{Baath} {et~al.}(1991){Baath}, {Padin}, {Woody}, {Rogers}, {Wright}, {Zensus}, {Kus}, {Backer}, {Booth}, {Carlstrom}, {Dickman}, {Emerson}, {Hirabayashi}, {Hodges}, {Inoue}, {Moran}, {Morimoto}, {Payne}, {Plambeck}, {Predmore}, \& {Ronnang}}]{baath1991}
{Baath}, L.~B., {Padin}, S., {Woody}, D., {et~al.} 1991, \aap, 241, L1

\bibitem[{{Baath} {et~al.}(1992){Baath}, {Rogers}, {Inoue}, {Padin}, {Wright}, {Zensus}, {Kus}, {Backer}, {Booth}, {Carlstrom}, {Dickman}, {Emerson}, {Hirabayashi}, {Hodges}, {Kobayashi}, {Lamb}, {Moran}, {Morimoto}, {Plambeck}, {Predmore}, {Ronnang}, \& {Woody}}]{baath1992}
{Baath}, L.~B., {Rogers}, A.~E.~E., {Inoue}, M., {et~al.} 1992, \aap, 257, 31

\bibitem[{{Belitsky} {et~al.}(2018){Belitsky}, {Lapkin, I.}, {Fredrixon, M.}, {Meledin, D.}, {Sundin, E.}, {Billade, B.}, {Ferm, S.-E.}, {Pavolotsky, A.}, {Rashid, H.}, {Strandberg, M.}, {Desmaris, V.}, {Ermakov, A.}, {Krause, S.}, {Olberg, M.}, {Aghdam, P.}, {Shafiee, S.}, {Bergman, P.}, {De Beck, E.}, {Olofsson, H.}, {Conway, J.}, {De Breuck, C.}, {Immer, K.}, {Yagoubov, P.}, {Montenegro-Montes, F. M.}, {Torstensson, K.}, {Pérez-Beaupuits, J.-P.}, {Klein, T.}, {Boland, W.}, {Baryshev, A. M.}, {Hesper, R.}, {Barkhof, J.}, {Adema, J.}, {Bekema, M. E.}, \& {Koops, A.}}]{Belitsky_2018}
{Belitsky}, V., {Lapkin, I.}, {Fredrixon, M.}, {et~al.} 2018, A\&A, 612, A23, \dodoi{10.1051/0004-6361/201731458}

\bibitem[{{Blackburn} {et~al.}(2019){Blackburn}, {Chan}, {Crew}, {Fish}, {Issaoun}, {Johnson}, {Wielgus}, {Akiyama}, {Barrett}, {Bouman}, {Cappallo}, {Chael}, {Janssen}, {Lonsdale}, \& {Doeleman}}]{blackburn2019a}
{Blackburn}, L., {Chan}, C.-k., {Crew}, G.~B., {et~al.} 2019, \apj, 882, 23, \dodoi{10.3847/1538-4357/ab328d}

\bibitem[{{Carpenter} {et~al.}(2023){Carpenter}, {Brogan}, {Iono}, \& {Mroczkowski}}]{ALMA_2030_WSU}
{Carpenter}, J., {Brogan}, C., {Iono}, D., \& {Mroczkowski}, T. 2023, in Physics and Chemistry of Star Formation: The Dynamical ISM Across Time and Spatial Scales, 304

\bibitem[{{Carter} {et~al.}(2012){Carter}, {Lazareff}, {Maier}, {Chenu}, {Fontana}, {Bortolotti}, {Boucher}, {Navarrini}, {Blanchet}, \& {Greve}}]{carter2012}
{Carter}, M., {Lazareff}, B., {Maier}, D., {et~al.} 2012, \aap, 538, A89, \dodoi{10.1051/0004-6361/201118452}

\bibitem[{{Chael} {et~al.}(2023){Chael}, {Issaoun}, {Pesce}, {Johnson}, {Ricarte}, {Fromm}, \& {Mizuno}}]{Chael_MFS_2023}
{Chael}, A., {Issaoun}, S., {Pesce}, D.~W., {et~al.} 2023, \apj, 945, 40, \dodoi{10.3847/1538-4357/acb7e4}

\bibitem[{{Chen} {et~al.}(2023){Chen}, {Asada}, {Matsushita}, {Raffin}, {Inoue}, {Ho}, {Han}, {Kubo}, {Norton}, {Patel}, {Nystrom}, {Huang}, {Martin-Cocher}, {Yi Koay}, {Romero-Ca{\~n}izales}, {Liu}, {Huang}, {Liu}, {Wei}, {Chang}, {Chilson}, {Oshiro}, {Jiang}, {Li}, {Bower}, {Shaw}, {Nishioka}, {Koch}, {Chen}, {Srinivasan}, {Rao}, {Snow}, {Jinchi}, {Han}, {Chang}, {Lu}, {Ogawa}, {Kimura}, {Hasegawa}, {Pu}, {Koyama}, {Nakamura}, {Bintley}, {Walther}, {Friberg}, {Dempsey}, {Sriharan}, {Srikanth}, {Doeleman}, {Brissenden}, {Algaba Marcos}, {Jeter}, {Kuo}, \& {Park}}]{Chen_2023}
{Chen}, M.-T., {Asada}, K., {Matsushita}, S., {et~al.} 2023, \pasp, 135, 095001, \dodoi{10.1088/1538-3873/acf072}

\bibitem[{Chenu {et~al.}(2007)Chenu, Carter, Maier, Bortolotti, Butin, Serres, Boucher, Mattiocco, \& Lazareff}]{chenu2007}
Chenu, J.~Y., Carter, M., Maier, D., {et~al.} 2007, in 2007 Joint 32nd International Conference on Infrared and Millimeter Waves and the 15th International Conference on Terahertz Electronics, 176--177

\bibitem[{{Chenu} {et~al.}(2016){Chenu}, {Navarrini}, {Bortolotti}, {Butin}, {Fontana}, {Mahieu}, { Maier}, {Mattiocco}, {Serres}, {Berton}, {Garnier}, {Moutote}, {Parioleau}, { Pissard}, \& {Reverdy}}]{chenu2016}
{Chenu}, J.-Y., {Navarrini}, A., {Bortolotti}, Y., {et~al.} 2016, IEEE Transactions on Terahertz Science and Technology, 6, 223, \dodoi{10.1109/TTHZ.2016.2525762}

\bibitem[{Crew {et~al.}(2023)Crew, Goddi, Matthews, Rottmann, Saez, \& MartÃ­-Vidal}]{crew2023}
Crew, G.~B., Goddi, C., Matthews, L.~D., {et~al.} 2023, Publications of the Astronomical Society of the Pacific, 135, 025002, \dodoi{10.1088/1538-3873/acb348}

\bibitem[{{Deller} {et~al.}(2011){Deller}, {Brisken}, {Phillips}, {Morgan}, {Alef}, {Cappallo}, {Middelberg}, {Romney}, {Rottmann}, {Tingay}, \& {Wayth}}]{deller2011}
{Deller}, A.~T., {Brisken}, W.~F., {Phillips}, C.~J., {et~al.} 2011, \pasp, 123, 275, \dodoi{10.1086/658907}

\bibitem[{{Doeleman} {et~al.}(2011){Doeleman}, {Mai}, {Rogers}, {Hartnett}, {Tobar}, \& {Nand}}]{Doeleman_2011}
{Doeleman}, S., {Mai}, T., {Rogers}, A. E.~E., {et~al.} 2011, \pasp, 123, 582, \dodoi{10.1086/660156}

\bibitem[{{Doeleman} {et~al.}(2019){Doeleman}, {Blackburn}, {Dexter}, {Gomez}, {Johnson}, {Palumbo}, {Weintroub}, {Farah}, {Fish}, {Loinard}, {Lonsdale}, {Narayanan}, {Patel}, {Pesce}, {Raymond}, {Tilanus}, {Wielgus}, {Akiyama}, {Bower}, {Broderick}, {Deane}, {Fromm}, {Gammie}, {Gold}, {Janssen}, {Kawashima}, {Krichbaum}, {Marrone}, {Matthews}, {Mizuno}, {Rezzolla}, {Roelofs}, {Ros}, {Savolainen}, {Yuan}, {Zhao}, {Blackburn}, {Doeleman}, {Dexter}, {Gomez}, {Johnson}, {Palumbo}, {Weintroub}, {Farah}, {Fish}, {Loinard}, {Lonsdale}, {Narayanan}, {Patel}, {Pesce}, {Raymond}, {Tilanus}, {Wielgus}, {Akiyama}, {Bower}, {Broderick}, {Deane}, {Fromm}, {Gammie}, {Gold}, {Janssen}, {Kawashima}, {Krichbaum}, {Marrone}, {Matthews}, {Mizuno}, {Rezzolla}, {Roelofs}, {Ros}, {Savolainen}, {Yuan}, \& {Zhao}}]{Doeleman_2019}
{Doeleman}, S., {Blackburn}, L., {Dexter}, J., {et~al.} 2019, in Bulletin of the American Astronomical Society, Vol.~51, 256

\bibitem[{{Doeleman} {et~al.}(2002){Doeleman}, {Phillips}, {Rogers}, {Attridge}, {Titus}, {Smythe}, {Cappallo}, {Buretta}, {Whitney}, {Krichbaum}, {Graham}, {Alef}, {Polatidis}, {Bach}, {Witzel}, {Zensus}, {Greve}, {Grewing}, {Freund}, {Strittmatter}, {Ziurys}, {Wilson}, {Fagg}, \& {Gay}}]{Doeleman_2002}
{Doeleman}, S.~S., {Phillips}, R.~B., {Rogers}, A.~E.~E., {et~al.} 2002, in Proceedings of the 6th EVN Symposium, 223

\bibitem[{{Doeleman} {et~al.}(2008){Doeleman}, {Weintroub}, {Rogers}, {Plambeck}, {Freund}, {Tilanus}, {Friberg}, {Ziurys}, {Moran}, {Corey}, {Young}, {Smythe}, {Titus}, {Marrone}, {Cappallo}, {Bock}, {Bower}, {Chamberlin}, {Davis}, {Krichbaum}, {Lamb}, {Maness}, {Niell}, {Roy}, {Strittmatter}, {Werthimer}, {Whitney}, \& {Woody}}]{doeleman2008}
{Doeleman}, S.~S., {Weintroub}, J., {Rogers}, A.~E.~E., {et~al.} 2008, \nat, 455, 78, \dodoi{10.1038/nature07245}

\bibitem[{{Doeleman} {et~al.}(2012){Doeleman}, {Fish}, {Schenck}, {Beaudoin}, {Blundell}, {Bower}, {Broderick}, {Chamberlin}, {Freund}, {Friberg}, {Gurwell}, {Ho}, {Honma}, {Inoue}, {Krichbaum}, {Lamb}, {Loeb}, {Lonsdale}, {Marrone}, {Moran}, {Oyama}, {Plambeck}, {Primiani}, {Rogers}, {Smythe}, {SooHoo}, {Strittmatter}, {Tilanus}, {Titus}, {Weintroub}, {Wright}, {Young}, \& {Ziurys}}]{Doeleman2012}
{Doeleman}, S.~S., {Fish}, V.~L., {Schenck}, D.~E., {et~al.} 2012, Science, 338, 355, \dodoi{10.1126/science.1224768}

\bibitem[{{Doeleman} {et~al.}(2023){Doeleman}, {Barrett}, {Blackburn}, {Bouman}, {Broderick}, {Chaves}, {Fish}, {Fitzpatrick}, {Freeman}, {Fuentes}, {G{\'o}mez}, {Haworth}, {Houston}, {Issaoun}, {Johnson}, {Kettenis}, {Loinard}, {Nagar}, {Narayanan}, {Oppenheimer}, {Palumbo}, {Patel}, {Pesce}, {Raymond}, {Roelofs}, {Srinivasan}, {Tiede}, {Weintroub}, \& {Wielgus}}]{ngeht_refarray}
{Doeleman}, S.~S., {Barrett}, J., {Blackburn}, L., {et~al.} 2023, Galaxies, 11, 107, \dodoi{10.3390/galaxies11050107}

\bibitem[{{Event Horizon Telescope Collaboration} {et~al.}(2019{\natexlab{a}}){Event Horizon Telescope Collaboration}, {Akiyama}, {Alberdi}, {Alef}, {Asada}, {Azulay}, {Baczko}, {Ball}, {Balokovi{\'c}}, {Barrett}, \& et~al.}]{paperi}
{Event Horizon Telescope Collaboration}, {Akiyama}, K., {Alberdi}, A., {et~al.} 2019{\natexlab{a}}, \apjl, 875, L1, \dodoi{10.3847/2041-8213/ab0ec7}

\bibitem[{{Event Horizon Telescope Collaboration} {et~al.}(2019{\natexlab{b}}){Event Horizon Telescope Collaboration}, {Akiyama}, {Alberdi}, {Alef}, {Asada}, {Azulay}, {Baczko}, {Ball}, {Balokovi{\'c}}, {Barrett}, \& et~al.}]{paperii}
---. 2019{\natexlab{b}}, \apjl, 875, L2, \dodoi{10.3847/2041-8213/ab0c96}

\bibitem[{{Event Horizon Telescope Collaboration} {et~al.}(2019{\natexlab{c}}){Event Horizon Telescope Collaboration}, {Akiyama}, {Alberdi}, {Alef}, {Asada}, {Azulay}, {Baczko}, {Ball}, {Balokovi{\'c}}, {Barrett}, \& et~al.}]{paperiv}
---. 2019{\natexlab{c}}, \apjl, 875, L4, \dodoi{10.3847/2041-8213/ab0e85}

\bibitem[{{Event Horizon Telescope Collaboration} {et~al.}(2019{\natexlab{d}}){Event Horizon Telescope Collaboration}, {Akiyama}, {Alberdi}, {Alef}, {Asada}, {Azulay}, {Baczko}, {Ball}, {Balokovi{\'c}}, {Barrett}, \& et~al.}]{paperiii}
---. 2019{\natexlab{d}}, \apjl, 875, L3, \dodoi{10.3847/2041-8213/ab0c57}

\bibitem[{{Event Horizon Telescope Collaboration} {et~al.}(2019{\natexlab{e}}){Event Horizon Telescope Collaboration}, {Akiyama}, {Alberdi}, {Alef}, {Asada}, {Azulay}, {Baczko}, {Ball}, {Balokovi{\'c}}, {Barrett}, {Bintley}, {Blackburn}, {Boland}, {Bouman}, {Bower}, {Bremer}, {Brinkerink}, {Brissenden}, {Britzen}, {Broderick}, {Broguiere}, {Bronzwaer}, {Byun}, {Carlstrom}, {Chael}, {Chan}, {Chatterjee}, {Chatterjee}, {Chen}, {Chen}, {Cho}, {Christian}, {Conway}, {Cordes}, {Crew}, {Cui}, {Davelaar}, {De Laurentis}, {Deane}, {Dempsey}, {Desvignes}, {Dexter}, {Doeleman}, {Eatough}, {Falcke}, {Fish}, {Fomalont}, {Fraga-Encinas}, {Friberg}, {Fromm}, {G{\'o}mez}, {Galison}, {Gammie}, {Garc{\'\i}a}, {Gentaz}, {Georgiev}, {Goddi}, {Gold}, {Gu}, {Gurwell}, {Hada}, {Hecht}, {Hesper}, {Ho}, {Ho}, {Honma}, {Huang}, {Huang}, {Hughes}, {Ikeda}, {Inoue}, {Issaoun}, {James}, {Jannuzi}, {Janssen}, {Jeter}, {Jiang}, {Johnson}, {Jorstad}, {Jung}, {Karami}, {Karuppusamy}, {Kawashima}, {Keating}, {Kettenis}, {Kim}, {Kim}, {Kim},
  {Kino}, {Koay}, {Koch}, {Koyama}, {Kramer}, {Kramer}, {Krichbaum}, {Kuo}, {Lauer}, {Lee}, {Li}, {Li}, {Lindqvist}, {Liu}, {Liuzzo}, {Lo}, {Lobanov}, {Loinard}, {Lonsdale}, {Lu}, {MacDonald}, {Mao}, {Markoff}, {Marrone}, {Marscher}, {Mart{\'\i}-Vidal}, {Matsushita}, {Matthews}, {Medeiros}, {Menten}, {Mizuno}, {Mizuno}, {Moran}, {Moriyama}, {Moscibrodzka}, {Mul{\ensuremath{\ddot{}}}ler}, {Nagai}, {Nagar}, {Nakamura}, {Narayan}, {Narayanan}, {Natarajan}, {Neri}, {Ni}, {Noutsos}, {Okino}, {Olivares}, {Oyama}, {{\"O}zel}, {Palumbo}, {Patel}, {Pen}, {Pesce}, {Pi{\'e}tu}, {Plambeck}, {PopStefanija}, {Porth}, {Prather}, {Preciado-L{\'o}pez}, {Psaltis}, {Pu}, {Ramakrishnan}, {Rao}, {Rawlings}, {Raymond}, {Rezzolla}, {Ripperda}, {Roelofs}, {Rogers}, {Ros}, {Rose}, {Roshanineshat}, {Rottmann}, {Roy}, {Ruszczyk}, {Ryan}, {Rygl}, {S{\'a}nchez}, {S{\'a}nchez-Arguelles}, {Sasada}, {Savolainen}, {Schloerb}, {Schuster}, {Shao}, {Shen}, {Small}, {Sohn}, {SooHoo}, {Tazaki}, {Tiede}, {Tilanus}, {Titus}, {Toma}, {Torne},
  {Trent}, {Trippe}, {Tsuda}, {van Bemmel}, {van Langevelde}, {van Rossum}, {Wagner}, {Wardle}, {Weintroub}, {Wex}, {Wharton}, {Wielgus}, {Wong}, {Wu}, {Young}, {Young}, {Younsi}, {Yuan}, {Yuan}, {Zensus}, {Zhao}, {Zhao}, {Zhu}, {Anczarski}, {Baganoff}, {Eckart}, {Farah}, {Haggard}, {Meyer-Zhao}, {Michalik}, {Nadolski}, {Neilsen}, {Nishioka}, {Nowak}, {Pradel}, {Primiani}, {Souccar}, {Vertatschitsch}, {Yamaguchi}, \& {Zhang}}]{M87PaperV}
---. 2019{\natexlab{e}}, \apjl, 875, L5, \dodoi{10.3847/2041-8213/ab0f43}

\bibitem[{{Event Horizon Telescope Collaboration} {et~al.}(2021){Event Horizon Telescope Collaboration}, {Akiyama}, {Algaba}, {Alberdi}, {Alef}, {Anantua}, {Asada}, {Azulay}, {Baczko}, {Ball}, {Balokovi{\'c}}, {Barrett}, {Benson}, {Bintley}, {Blackburn}, {Blundell}, {Boland}, {Bouman}, {Bower}, {Boyce}, {Bremer}, {Brinkerink}, {Brissenden}, {Britzen}, {Broderick}, {Broguiere}, {Bronzwaer}, {Byun}, {Carlstrom}, {Chael}, {Chan}, {Chatterjee}, {Chatterjee}, {Chen}, {Chen}, {Chesler}, {Cho}, {Christian}, {Conway}, {Cordes}, {Crawford}, {Crew}, {Cruz-Osorio}, {Cui}, {Davelaar}, {De Laurentis}, {Deane}, {Dempsey}, {Desvignes}, {Dexter}, {Doeleman}, {Eatough}, {Falcke}, {Farah}, {Fish}, {Fomalont}, {Ford}, {Fraga-Encinas}, {Friberg}, {Fromm}, {Fuentes}, {Galison}, {Gammie}, {Garc{\'\i}a}, {Gelles}, {Gentaz}, {Georgiev}, {Goddi}, {Gold}, {G{\'o}mez}, {G{\'o}mez-Ruiz}, {Gu}, {Gurwell}, {Hada}, {Haggard}, {Hecht}, {Hesper}, {Himwich}, {Ho}, {Ho}, {Honma}, {Huang}, {Huang}, {Hughes}, {Ikeda}, {Inoue}, {Issaoun},
  {James}, {Jannuzi}, {Janssen}, {Jeter}, {Jiang}, {Jimenez-Rosales}, {Johnson}, {Jorstad}, {Jung}, {Karami}, {Karuppusamy}, {Kawashima}, {Keating}, {Kettenis}, {Kim}, {Kim}, {Kim}, {Kim}, {Kino}, {Koay}, {Kofuji}, {Koch}, {Koyama}, {Kramer}, {Kramer}, {Krichbaum}, {Kuo}, {Lauer}, {Lee}, {Levis}, {Li}, {Li}, {Lindqvist}, {Lico}, {Lindahl}, {Liu}, {Liu}, {Liuzzo}, {Lo}, {Lobanov}, {Loinard}, {Lonsdale}, {Lu}, {MacDonald}, {Mao}, {Marchili}, {Markoff}, {Marrone}, {Marscher}, {Mart{\'\i}-Vidal}, {Matsushita}, {Matthews}, {Medeiros}, {Menten}, {Mizuno}, {Mizuno}, {Moran}, {Moriyama}, {Moscibrodzka}, {M{\"u}ller}, {Musoke}, {Mus Mej{\'\i}as}, {Michalik}, {Nadolski}, {Nagai}, {Nagar}, {Nakamura}, {Narayan}, {Narayanan}, {Natarajan}, {Nathanail}, {Neilsen}, {Neri}, {Ni}, {Noutsos}, {Nowak}, {Okino}, {Olivares}, {Ortiz-Le{\'o}n}, {Oyama}, {{\"O}zel}, {Palumbo}, {Park}, {Patel}, {Pen}, {Pesce}, {Pi{\'e}tu}, {Plambeck}, {PopStefanija}, {Porth}, {P{\"o}tzl}, {Prather}, {Preciado-L{\'o}pez}, {Psaltis}, {Pu},
  {Ramakrishnan}, {Rao}, {Rawlings}, {Raymond}, {Rezzolla}, {Ricarte}, {Ripperda}, {Roelofs}, {Rogers}, {Ros}, {Rose}, {Roshanineshat}, {Rottmann}, {Roy}, {Ruszczyk}, {Rygl}, {S{\'a}nchez}, {S{\'a}nchez-Arguelles}, {Sasada}, {Savolainen}, {Schloerb}, {Schuster}, {Shao}, {Shen}, {Small}, {Sohn}, {SooHoo}, {Sun}, {Tazaki}, {Tetarenko}, {Tiede}, {Tilanus}, {Titus}, {Toma}, {Torne}, {Trent}, {Traianou}, {Trippe}, {van Bemmel}, {van Langevelde}, {van Rossum}, {Wagner}, {Ward-Thompson}, {Wardle}, {Weintroub}, {Wex}, {Wharton}, {Wielgus}, {Wong}, {Wu}, {Yoon}, {Young}, {Young}, {Younsi}, {Yuan}, {Yuan}, {Zensus}, {Zhao}, \& {Zhao}}]{paperviii}
{Event Horizon Telescope Collaboration}, {Akiyama}, K., {Algaba}, J.~C., {et~al.} 2021, The Astrophysical Journal Letters, 910, L13, \dodoi{10.3847/2041-8213/abe4de}

\bibitem[{{Event Horizon Telescope Collaboration} {et~al.}(2022{\natexlab{a}}){Event Horizon Telescope Collaboration}, {Akiyama}, {Alberdi}, {Alef}, {Algaba}, {Anantua}, {Asada}, {Azulay}, {Bach}, {Baczko}, {Ball}, {Balokovi{\'c}}, {Barrett}, {Baub{\"o}ck}, {Benson}, {Bintley}, {Blackburn}, {Blundell}, {Bouman}, {Bower}, {Boyce}, {Bremer}, {Brinkerink}, {Brissenden}, {Britzen}, {Broderick}, {Broguiere}, {Bronzwaer}, {Bustamante}, {Byun}, {Carlstrom}, {Ceccobello}, {Chael}, {Chan}, {Chatterjee}, {Chatterjee}, {Chen}, {Chen}, {Cheng}, {Cho}, {Christian}, {Conroy}, {Conway}, {Cordes}, {Crawford}, {Crew}, {Cruz-Osorio}, {Cui}, {Davelaar}, {Laurentis}, {Deane}, {Dempsey}, {Desvignes}, {Dexter}, {Dhruv}, {Doeleman}, {Dougal}, {Dzib}, {Eatough}, {Emami}, {Falcke}, {Farah}, {Fish}, {Fomalont}, {Ford}, {Fraga-Encinas}, {Freeman}, {Friberg}, {Fromm}, {Fuentes}, {Galison}, {Gammie}, {Garc{\'\i}a}, {Gentaz}, {Georgiev}, {Goddi}, {Gold}, {G{\'o}mez-Ruiz}, {G{\'o}mez}, {Gu}, {Gurwell}, {Hada}, {Haggard}, {Haworth},
  {Hecht}, {Hesper}, {Heumann}, {Ho}, {Ho}, {Honma}, {Huang}, {Huang}, {Hughes}, {Ikeda}, {Impellizzeri}, {Inoue}, {Issaoun}, {James}, {Jannuzi}, {Janssen}, {Jeter}, {Jiang}, {Jim{\'e}nez-Rosales}, {Johnson}, {Jorstad}, {Joshi}, {Jung}, {Karami}, {Karuppusamy}, {Kawashima}, {Keating}, {Kettenis}, {Kim}, {Kim}, {Kim}, {Kim}, {Kino}, {Koay}, {Kocherlakota}, {Kofuji}, {Koch}, {Koyama}, {Kramer}, {Kramer}, {Krichbaum}, {Kuo}, {Bella}, {Lauer}, {Lee}, {Lee}, {Leung}, {Levis}, {Li}, {Lico}, {Lindahl}, {Lindqvist}, {Lisakov}, {Liu}, {Liu}, {Liuzzo}, {Lo}, {Lobanov}, {Loinard}, {Lonsdale}, {Lu}, {Mao}, {Marchili}, {Markoff}, {Marrone}, {Marscher}, {Mart{\'\i}-Vidal}, {Matsushita}, {Matthews}, {Medeiros}, {Menten}, {Michalik}, {Mizuno}, {Mizuno}, {Moran}, {Moriyama}, {Moscibrodzka}, {M{\"u}ller}, {Mus}, {Musoke}, {Myserlis}, {Nadolski}, {Nagai}, {Nagar}, {Nakamura}, {Narayan}, {Narayanan}, {Natarajan}, {Nathanail}, {Fuentes}, {Neilsen}, {Neri}, {Ni}, {Noutsos}, {Nowak}, {Oh}, {Okino}, {Olivares}, {Ortiz-Le{\'o}n},
  {Oyama}, {{\"O}zel}, {Palumbo}, {Paraschos}, {Park}, {Parsons}, {Patel}, {Pen}, {Pesce}, {Pi{\'e}tu}, {Plambeck}, {PopStefanija}, {Porth}, {P{\"o}tzl}, {Prather}, {Preciado-L{\'o}pez}, {Psaltis}, {Pu}, {Ramakrishnan}, {Rao}, {Rawlings}, {Raymond}, {Rezzolla}, {Ricarte}, {Ripperda}, {Roelofs}, {Rogers}, {Ros}, {Romero-Ca{\~n}izales}, {Roshanineshat}, {Rottmann}, {Roy}, {Ruiz}, {Ruszczyk}, {Rygl}, {S{\'a}nchez}, {S{\'a}nchez-Arg{\"u}elles}, {S{\'a}nchez-Portal}, {Sasada}, {Satapathy}, {Savolainen}, {Schloerb}, {Schonfeld}, {Schuster}, {Shao}, {Shen}, {Small}, {Sohn}, {SooHoo}, {Souccar}, {Sun}, {Tazaki}, {Tetarenko}, {Tiede}, {Tilanus}, {Titus}, {Torne}, {Traianou}, {Trent}, {Trippe}, {Turk}, {van Bemmel}, {van Langevelde}, {van Rossum}, {Vos}, {Wagner}, {Ward-Thompson}, {Wardle}, {Weintroub}, {Wex}, {Wharton}, {Wielgus}, {Wiik}, {Witzel}, {Wondrak}, {Wong}, {Wu}, {Yamaguchi}, {Yoon}, {Young}, {Young}, {Younsi}, {Yuan}, {Yuan}, {Zensus}, {Zhang}, {Zhao}, {Zhao}, {Agurto}, {Allardi}, {Amestica}, {Araneda},
  {Arriagada}, {Berghuis}, {Bertarini}, {Berthold}, {Blanchard}, {Brown}, {C{\'a}rdenas}, {Cantzler}, {Caro}, {Castillo-Dom{\'\i}nguez}, {Chan}, {Chang}, {Chang}, {Chang}, {Chang}, {Chen}, {Chilson}, {Chuter}, {Ciechanowicz}, {Colin-Beltran}, {Coulson}, {Crowley}, {Degenaar}, {Dornbusch}, {Dur{\'a}n}, {Everett}, {Faber}, {Forster}, {Fuchs}, {Gale}, {Geertsema}, {Gonz{\'a}lez}, {Graham}, {Gueth}, {Halverson}, {Han}, {Han}, {Hasegawa}, {Hern{\'a}ndez-Rebollar}, {Herrera}, {Herrero-Illana}, {Heyminck}, {Hirota}, {Hoge}, {Hostler Schimpf}, {Howie}, {Huang}, {Jiang}, {Jinchi}, {John}, {Kimura}, {Klein}, {Kubo}, {Kuroda}, {Kwon}, {Lacasse}, {Laing}, {Leitch}, {Li}, {Liu}, {Liu}, {Lin}, {Lu}, {Mac-Auliffe}, {Martin-Cocher}, {Matulonis}, {Maute}, {Messias}, {Meyer-Zhao}, {Monta{\~n}a}, {Montenegro-Montes}, {Montgomerie}, {Moreno Nolasco}, {Muders}, {Nishioka}, {Norton}, {Nystrom}, {Ogawa}, {Olivares}, {Oshiro}, {P{\'e}rez-Beaupuits}, {Parra}, {Phillips}, {Poirier}, {Pradel}, {Qiu}, {Raffin}, {Rahlin}, {Ram{\'\i}rez},
  {Ressler}, {Reynolds}, {Rodr{\'\i}guez-Montoya}, {Saez-Madain}, {Santana}, {Shaw}, {Shirkey}, {Silva}, {Snow}, {Sousa}, {Sridharan}, {Stahm}, {Stark}, {Test}, {Torstensson}, {Venegas}, {Walther}, {Wei}, {White}, {Wieching}, {Wijnands}, {Wouterloot}, {Yu}, {Yu (于威)}, \& {Zeballos}}]{sgr_paper1}
{Event Horizon Telescope Collaboration}, {Akiyama}, K., {Alberdi}, A., {et~al.} 2022{\natexlab{a}}, \apjl, 930, L12, \dodoi{10.3847/2041-8213/ac6674}

\bibitem[{{Event Horizon Telescope Collaboration} {et~al.}(2022{\natexlab{b}}){Event Horizon Telescope Collaboration}, {Akiyama}, {Alberdi}, {Alef}, {Algaba}, {Anantua}, {Asada}, {Azulay}, {Bach}, {Baczko}, {Ball}, {Balokovi{\'c}}, {Barrett}, {Baub{\"o}ck}, {Benson}, {Bintley}, {Blackburn}, {Blundell}, {Bouman}, {Bower}, {Boyce}, {Bremer}, {Brinkerink}, {Brissenden}, {Britzen}, {Broderick}, {Broguiere}, {Bronzwaer}, {Bustamante}, {Byun}, {Carlstrom}, {Ceccobello}, {Chael}, {Chan}, {Chatterjee}, {Chatterjee}, {Chen}, {Chen}, {Cheng}, {Cho}, {Christian}, {Conroy}, {Conway}, {Cordes}, {Crawford}, {Crew}, {Cruz-Osorio}, {Cui}, {Davelaar}, {De Laurentis}, {Deane}, {Dempsey}, {Desvignes}, {Dexter}, {Dhruv}, {Doeleman}, {Dougal}, {Dzib}, {Eatough}, {Emami}, {Falcke}, {Farah}, {Fish}, {Fomalont}, {Ford}, {Fraga-Encinas}, {Freeman}, {Friberg}, {Fromm}, {Fuentes}, {Galison}, {Gammie}, {Garc{\'\i}a}, {Gentaz}, {Georgiev}, {Goddi}, {Gold}, {G{\'o}mez-Ruiz}, {G{\'o}mez}, {Gu}, {Gurwell}, {Hada}, {Haggard}, {Haworth},
  {Hecht}, {Hesper}, {Heumann}, {Ho}, {Ho}, {Honma}, {Huang}, {Huang}, {Hughes}, {Ikeda}, {Violette Impellizzeri}, {Inoue}, {Issaoun}, {James}, {Jannuzi}, {Janssen}, {Jeter}, {Jiang}, {Jim{\'e}nez-Rosales}, {Johnson}, {Jorstad}, {Joshi}, {Jung}, {Karami}, {Karuppusamy}, {Kawashima}, {Keating}, {Kettenis}, {Kim}, {Kim}, {Kim}, {Kim}, {Kino}, {Koay}, {Kocherlakota}, {Kofuji}, {Koch}, {Koyama}, {Kramer}, {Kramer}, {Krichbaum}, {Kuo}, {La Bella}, {Lauer}, {Lee}, {Lee}, {Leung}, {Levis}, {Li}, {Lico}, {Lindahl}, {Lindqvist}, {Lisakov}, {Liu}, {Liu}, {Liuzzo}, {Lo}, {Lobanov}, {Loinard}, {Lonsdale}, {Lu}, {Mao}, {Marchili}, {Markoff}, {Marrone}, {Marscher}, {Mart{\'\i}-Vidal}, {Matsushita}, {Matthews}, {Medeiros}, {Menten}, {Michalik}, {Mizuno}, {Mizuno}, {Moran}, {Moriyama}, {Moscibrodzka}, {M{\"u}ller}, {Mus}, {Musoke}, {Myserlis}, {Nadolski}, {Nagai}, {Nagar}, {Nakamura}, {Narayan}, {Narayanan}, {Natarajan}, {Nathanail}, {Navarro Fuentes}, {Neilsen}, {Neri}, {Ni}, {Noutsos}, {Nowak}, {Oh}, {Okino}, {Olivares},
  {Ortiz-Le{\'o}n}, {Oyama}, {{\"O}zel}, {Palumbo}, {Filippos Paraschos}, {Park}, {Parsons}, {Patel}, {Pen}, {Pesce}, {Pi{\'e}tu}, {Plambeck}, {PopStefanija}, {Porth}, {P{\"o}tzl}, {Prather}, {Preciado-L{\'o}pez}, {Psaltis}, {Pu}, {Ramakrishnan}, {Rao}, {Rawlings}, {Raymond}, {Rezzolla}, {Ricarte}, {Ripperda}, {Roelofs}, {Rogers}, {Ros}, {Romero-Ca{\~n}izales}, {Roshanineshat}, {Rottmann}, {Roy}, {Ruiz}, {Ruszczyk}, {Rygl}, {S{\'a}nchez}, {S{\'a}nchez-Arg{\"u}elles}, {S{\'a}nchez-Portal}, {Sasada}, {Satapathy}, {Savolainen}, {Schloerb}, {Schonfeld}, {Schuster}, {Shao}, {Shen}, {Small}, {Sohn}, {SooHoo}, {Souccar}, {Sun}, {Tazaki}, {Tetarenko}, {Tiede}, {Tilanus}, {Titus}, {Torne}, {Traianou}, {Trent}, {Trippe}, {Turk}, {van Bemmel}, {van Langevelde}, {van Rossum}, {Vos}, {Wagner}, {Ward-Thompson}, {Wardle}, {Weintroub}, {Wex}, {Wharton}, {Wielgus}, {Wiik}, {Witzel}, {Wondrak}, {Wong}, {Wu}, {Yamaguchi}, {Yoon}, {Young}, {Young}, {Younsi}, {Yuan}, {Yuan}, {Zensus}, {Zhang}, {Zhao}, {Zhao}, {Chan}, {Qiu},
  {Ressler}, \& {White}}]{SgrAPaperV}
---. 2022{\natexlab{b}}, \apjl, 930, L16, \dodoi{10.3847/2041-8213/ac6672}

\bibitem[{{Event Horizon Telescope Collaboration} {et~al.}(2024){Event Horizon Telescope Collaboration}, {Akiyama}, {Alberdi}, {Alef}, {Algaba}, {Anantua}, {Asada}, {Azulay}, {Bach}, {Baczko}, {Ball}, {Balokovi{\'c}}, {Bandyopadhyay}, {Barrett}, {Baub{\"o}ck}, {Benson}, {Bintley}, {Blackburn}, {Blundell}, {Bouman}, {Bower}, {Boyce}, {Bremer}, {Brissenden}, {Britzen}, {Broderick}, {Broguiere}, {Bronzwaer}, {Bustamante}, {Carlstrom}, {Chael}, {Chan}, {Chang}, {Chatterjee}, {Chatterjee}, {Chen}, {Chen}, {Cheng}, {Cho}, {Christian}, {Conroy}, {Conway}, {Crawford}, {Crew}, {Cruz-Osorio}, {Cui}, {Dahale}, {Davelaar}, {De Laurentis}, {Deane}, {Dempsey}, {Desvignes}, {Dexter}, {Dhruv}, {Dihingia}, {Doeleman}, {Dzib}, {Eatough}, {Emami}, {Falcke}, {Farah}, {Fish}, {Fomalont}, {Ford}, {Foschi}, {Fraga-Encinas}, {Freeman}, {Friberg}, {Fromm}, {Fuentes}, {Galison}, {Gammie}, {Garc{\'\i}a}, {Gentaz}, {Georgiev}, {Goddi}, {Gold}, {G{\'o}mez-Ruiz}, {G{\'o}mez}, {Gu}, {Gurwell}, {Hada}, {Haggard}, {Hesper}, {Heumann}, {Ho},
  {Ho}, {Honma}, {Huang}, {Huang}, {Hughes}, {Ikeda}, {Violette Impellizzeri}, {Inoue}, {Issaoun}, {James}, {Jannuzi}, {Janssen}, {Jeter}, {Jiang}, {Jim{\'e}nez-Rosales}, {Johnson}, {Jorstad}, {Jones}, {Joshi}, {Jung}, {Karuppusamy}, {Kawashima}, {Keating}, {Kettenis}, {Kim}, {Kim}, {Kim}, {Kim}, {Kino}, {Koay}, {Kocherlakota}, {Kofuji}, {Koch}, {Koyama}, {Kramer}, {Kramer}, {Kramer}, {Krichbaum}, {Kuo}, {La Bella}, {Lee}, {Levis}, {Li}, {Lico}, {Lindahl}, {Lindqvist}, {Lisakov}, {Liu}, {Liu}, {Liuzzo}, {Lo}, {Lobanov}, {Loinard}, {Lonsdale}, {Lowitz}, {Lu}, {MacDonald}, {Mao}, {Marchili}, {Markoff}, {Marrone}, {Marscher}, {Mart{\'\i}-Vidal}, {Matsushita}, {Matthews}, {Medeiros}, {Menten}, {Mizuno}, {Mizuno}, {Montgomery}, {Moran}, {Moriyama}, {Moscibrodzka}, {Mulaudzi}, {M{\"u}ller}, {M{\"u}ller}, {Mus}, {Musoke}, {Myserlis}, {Nagai}, {Nagar}, {Nakamura}, {Narayanan}, {Natarajan}, {Nathanail}, {Fuentes}, {Neilsen}, {Ni}, {Nowak}, {Oh}, {Okino}, {Olivares}, {Oyama}, {{\"O}zel}, {Palumbo}, {Paraschos}, {Park},
  {Parsons}, {Patel}, {Pen}, {Pesce}, {Pi{\'e}tu}, {PopStefanija}, {Porth}, {Prather}, {Psaltis}, {Pu}, {Ramakrishnan}, {Rao}, {Rawlings}, {Raymond}, {Rezzolla}, {Ricarte}, {Ripperda}, {Roelofs}, {Romero-Ca{\~n}izales}, {Ros}, {Roshanineshat}, {Rottmann}, {Roy}, {Ruiz}, {Ruszczyk}, {Rygl}, {S{\'a}nchez}, {S{\'a}nchez-Arg{\"u}elles}, {S{\'a}nchez-Portal}, {Sasada}, {Satapathy}, {Savolainen}, {Schloerb}, {Schonfeld}, {Schuster}, {Shao}, {Shen}, {Small}, {Sohn}, {SooHoo}, {Salas}, {Souccar}, {Stanway}, {Sun}, {Tazaki}, {Tetarenko}, {Tiede}, {Tilanus}, {Titus}, {Toma}, {Torne}, {Toscano}, {Traianou}, {Trent}, {Trippe}, {Turk}, {van Bemmel}, {van Langevelde}, {van Rossum}, {Vos}, {Wagner}, {Ward-Thompson}, {Wardle}, {Washington}, {Weintroub}, {Wharton}, {Wielgus}, {Wiik}, {Witzel}, {Wondrak}, {Wong}, {Wu}, {Yadlapalli}, {Yamaguchi}, {Yfantis}, {Yoon}, {Young}, {Younsi}, {Yu}, {Yuan}, {Yuan}, {Anton Zensus}, {Zhang}, {Zhao}, {Zhao}, {Allardi}, {Chang}, {Chang}, {Chang}, {Chen}, {Chilson}, {Faber}, {Gale}, {Han},
  {Han}, {Hasegawa}, {Hern{\'a}ndez-Rebollar}, {Huang}, {Jiang}, {Jinchi}, {Kimura}, {Kubo}, {Li}, {Lin}, {Liu}, {Liu}, {Lu}, {Martin-Cocher}, {Meyer-Zhao}, {Monta{\~n}a}, {Moraghan}, {Moreno-Nolasco}, {Nishioka}, {Norton}, {Nystrom}, {Ogawa}, {Oshiro}, {Pradel}, {Principe}, {Raffin}, {Rodr{\'\i}guez-Montoya}, {Shaw}, {Snow}, {Sridharan}, {Srinivasan}, {Wei}, \& {Yu}}]{M87_2018_aa}
---. 2024, \aap, 681, A79, \dodoi{10.1051/0004-6361/202347932}

\bibitem[{{Goddi} {et~al.}(2019){Goddi}, {Mart{\'\i}-Vidal}, {Messias}, {Crew}, {Herrero-Illana}, {Impellizzeri}, {Rottmann}, {Wagner}, {Fomalont}, {Matthews}, {Petry}, {Phillips}, {Tilanus}, {Villard}, {Blackburn}, {Janssen}, \& {Wielgus}}]{goddi2019}
{Goddi}, C., {Mart{\'\i}-Vidal}, I., {Messias}, H., {et~al.} 2019, \pasp, 131, 075003, \dodoi{10.1088/1538-3873/ab136a}

\bibitem[{{Greve} \& {Bremer}(2010)}]{2010ASSL..364.....G}
{Greve}, A., \& {Bremer}, M. 2010, {Thermal Design and Thermal Behaviour of Radio Telescopes and their Enclosures}, Vol. 364, \dodoi{10.1007/978-3-642-03866-2}

\bibitem[{{Greve} {et~al.}(1995){Greve}, {Torres}, {Wink}, {Grewing}, {Wild}, {Alcolea}, {Barcia}, {Colomer}, {de Vincente}, {Gomez-Gonzalez}, {Lopez-Fernandez}, {Graham}, {Krichbaum}, {Schwartz}, {Standke}, {Witzel}, \& {Baudry}}]{Greve_1995}
{Greve}, A., {Torres}, M., {Wink}, J.~E., {et~al.} 1995, \aap, 299, L33

\bibitem[{{Greve} {et~al.}(2002){Greve}, {K{\"o}n{\"o}nen}, {Graham}, {Wiik}, {Krichbaum}, {Conway}, {Rantakyr{\"o}}, {Urpo}, {Grewing}, {Booth}, {Zensus}, {John}, {Navarro}, {Mujunen}, {Ritakari}, {Peltonen}, {Sj{\"o}man}, {Oinaskallio}, \& {Berton}}]{Greve_2002}
{Greve}, A., {K{\"o}n{\"o}nen}, P., {Graham}, D.~A., {et~al.} 2002, \aap, 390, L19, \dodoi{10.1051/0004-6361:20020893}

\bibitem[{{Hada} {et~al.}(2011){Hada}, {Doi}, {Kino}, {Nagai}, {Hagiwara}, \& {Kawaguchi}}]{Hada_2011}
{Hada}, K., {Doi}, A., {Kino}, M., {et~al.} 2011, \nat, 477, 185, \dodoi{10.1038/nature10387}

\bibitem[{{Han} {et~al.}(2018){Han}, {Chen}, {Huang}, {Kubo}, {Chang}, {Chang}, {Wei}, {Huang}, {Chen}, {Raffin}, {Liu}, {Ho}, {Inoue}, {Matsushita}, {Asada}, {Norton}, {Chilson}, {Srinivasan}, {Liu}, {Li}, {Bintley}, {Walther}, {Friberg}, {Dempsey}, {Ogawa}, {Kimura}, {Hasagawa}, \& {Srikanth}}]{chihchiang2018}
{Han}, C.-C., {Chen}, M.-T., {Huang}, Y.-D., {et~al.} 2018, in Society of Photo-Optical Instrumentation Engineers (SPIE) Conference Series, Vol. 10708, Millimeter, Submillimeter, and Far-Infrared Detectors and Instrumentation for Astronomy IX, ed. J.~{Zmuidzinas} \& J.-R. {Gao}, 1070835

\bibitem[{{Inoue} {et~al.}(2014){Inoue}, {Algaba-Marcos}, {Asada}, {Blundell}, {Brisken}, {Burgos}, {Chang}, {Chen}, {Doeleman}, {Fish}, {Grimes}, {Han}, {Hirashita}, {Ho}, {Hsieh}, {Huang}, {Jiang}, {Keto}, {Koch}, {Kubo}, {Kuo}, {Liu}, {Martin-Cocher}, {Matsushita}, {Meyer-Zhao}, {Nakamura}, {Napier}, {Nishioka}, {Nystrom}, {Paine}, {Patel}, {Pradel}, {Pu}, {Raffin}, {Shen}, {Snow}, {Srinivasan}, \& {Wei}}]{Inoue_2014}
{Inoue}, M., {Algaba-Marcos}, J.~C., {Asada}, K., {et~al.} 2014, Radio Science, 49, 564, \dodoi{10.1002/2014RS005450}

\bibitem[{{Issaoun} {et~al.}(2019){Issaoun}, {Johnson}, {Blackburn}, {Brinkerink}, {Mo{\'s}cibrodzka}, {Chael}, {Goddi}, {Mart{\'\i}-Vidal}, {Wagner}, {Doeleman}, {Falcke}, {Krichbaum}, {Akiyama}, {Bach}, {Bouman}, {Bower}, {Broderick}, {Cho}, {Crew}, {Dexter}, {Fish}, {Gold}, {G{\'o}mez}, {Hada}, {Hern{\'a}ndez-G{\'o}mez}, {Jan{\ss}en}, {Kino}, {Kramer}, {Loinard}, {Lu}, {Markoff}, {Marrone}, {Matthews}, {Moran}, {M{\"u}ller}, {Roelofs}, {Ros}, {Rottmann}, {Sanchez}, {Tilanus}, {de Vicente}, {Wielgus}, {Zensus}, \& {Zhao}}]{issaoun2019}
{Issaoun}, S., {Johnson}, M.~D., {Blackburn}, L., {et~al.} 2019, \apj, 871, 30, \dodoi{10.3847/1538-4357/aaf732}

\bibitem[{{Issaoun} {et~al.}(2022){Issaoun}, {Wielgus}, {Jorstad}, {Krichbaum}, {Blackburn}, {Janssen}, {Chan}, {Pesce}, {G{\'o}mez}, {Akiyama}, {Mo{\'s}cibrodzka}, {Mart{\'\i}-Vidal}, {Chael}, {Lico}, {Liu}, {Ramakrishnan}, {Lisakov}, {Fuentes}, {Zhao}, {Moriyama}, {Broderick}, {Tiede}, {MacDonald}, {Mizuno}, {Traianou}, {Loinard}, {Davelaar}, {Gurwell}, {Lu}, {Alberdi}, {Alef}, {Algaba}, {Anantua}, {Asada}, {Azulay}, {Bach}, {Baczko}, {Ball}, {Balokovi{\'c}}, {Barrett}, {Baub{\"o}ck}, {Benson}, {Bintley}, {Blundell}, {Boland}, {Bouman}, {Bower}, {Boyce}, {Bremer}, {Brinkerink}, {Brissenden}, {Britzen}, {Broguiere}, {Bronzwaer}, {Bustamante}, {Byun}, {Carlstrom}, {Ceccobello}, {Chatterjee}, {Chatterjee}, {Chen}, {Chen}, {Cho}, {Christian}, {Conroy}, {Conway}, {Cordes}, {Crawford}, {Crew}, {Cruz-Osorio}, {Cui}, {De Laurentis}, {Deane}, {Dempsey}, {Desvignes}, {Dexter}, {Doeleman}, {Dhruv}, {Dzib Quijano}, {Eatough}, {Emami}, {Falcke}, {Farah}, {Fish}, {Fomalont}, {Ford}, {Fraga-Encinas}, {Freeman}, {Friberg},
  {Fromm}, {Galison}, {Gammie}, {Garc{\'\i}a}, {Gentaz}, {Georgiev}, {Goddi}, {Gold}, {G{\'o}mez-Ruiz}, {Gu}, {Hada}, {Haggard}, {Hecht}, {Hesper}, {Ho}, {Ho}, {Honma}, {Huang}, {Huang}, {Hughes}, {Ikeda}, {Impellizzeri}, {Inoue}, {James}, {Jannuzi}, {Jeter}, {Jiang}, {Jimenez-Rosales}, {Johnson}, {Joshi}, {Jung}, {Karami}, {Karuppusamy}, {Kawashima}, {Keating}, {Kettenis}, {Kim}, {Kim}, {Kim}, {Kim}, {Kino}, {Koay}, {Kocherlakota}, {Kofuji}, {Koch}, {Koyama}, {Kramer}, {Kramer}, {Kuo}, {La Bella}, {Lauer}, {Lee}, {Lee}, {Leung}, {Levis}, {Li}, {Lico}, {Lindahl}, {Lindqvist}, {Liu}, {Liuzzo}, {Lo}, {Lobanov}, {Lonsdale}, {Mao}, {Marchili}, {Markoff}, {Marrone}, {Marscher}, {Matsushita}, {Matthews}, {Medeiros}, {Menten}, {Michalik}, {Mizuno}, {Mizuno}, {Moran}, {M{\"u}ller}, {Mus}, {Musoke}, {Myserlis}, {Nadolski}, {Nagai}, {Nagar}, {Nakamura}, {Narayan}, {Narayanan}, {Natarajan}, {Nathanail}, {Neilsen}, {Neri}, {Ni}, {Noutsos}, {Nowak}, {Oh}, {Okino}, {Olivares}, {Ortiz-Le{\'o}n}, {Oyama}, {{\"O}zel},
  {Palumbo}, {Paraschos}, {Park}, {Parsons}, {Patel}, {Pen}, {Pi{\'e}tu}, {Plambeck}, {PopStefanija}, {Porth}, {P{\"o}tzl}, {Prather}, {Preciado-L{\'o}pez}, {Psaltis}, {Pu}, {Rao}, {Rawlings}, {Raymond}, {Rezzolla}, {Ricarte}, {Ripperda}, {Roelofs}, {Rogers}, {Ros}, {Romero-Canizales}, {Roshanineshat}, {Rottmann}, {Roy}, {Ruiz}, {Ruszczyk}, {Rygl}, {S{\'a}nchez}, {S{\'a}nchez-Arguelles}, {Sanchez-Portal}, {Sasada}, {Satapathy}, {Savolainen}, {Schloerb}, {Schuster}, {Shao}, {Shen}, {Small}, {Sohn}, {SooHoo}, {Souccar}, {Sun}, {Tazaki}, {Tetarenko}, {Tiede}, {Tilanus}, {Titus}, {Torne}, {Trent}, {Trippe}, {van Bemmel}, {van Langevelde}, {van Rossum}, {Vos}, {Wagner}, {Ward-Thompson}, {Wardle}, {Weintroub}, {Wex}, {Wharton}, {Wiik}, {Witzel}, {Wondrak}, {Wong}, {Wu}, {Yamaguchi}, {Yoon}, {Young}, {Young}, {Younsi}, {Yuan}, {Yuan}, {Zensus}, {Zhang}, \& {Zhao}}]{Issaoun2022}
{Issaoun}, S., {Wielgus}, M., {Jorstad}, S., {et~al.} 2022, \apj, 934, 145, \dodoi{10.3847/1538-4357/ac7a40}

\bibitem[{Janssen {et~al.}(2021)Janssen, Falcke, Kadler, Ros, Wielgus, Akiyama, Balokovi{\'{c}}, Blackburn, Bouman, Chael, Chan, Chatterjee, Davelaar, Edwards, Fromm, G{\'{o}}mez, Goddi, Issaoun, Johnson, Kim, Koay, Krichbaum, Liu, Liuzzo, Markoff, Markowitz, Marrone, Mizuno, M{\"{u}}ller, Ni, Pesce, Ramakrishnan, Roelofs, Rygl, van Bemmel, Alberdi, Alef, Algaba, Anantua, Asada, Azulay, Baczko, Ball, Barrett, Benson, Bintley, Blundell, Boland, Bower, Boyce, Bremer, Brinkerink, Brissenden, Britzen, Broderick, Broguiere, Bronzwaer, Byun, Carlstrom, Chatterjee, Chen, Chen, Chesler, Cho, Christian, Conway, Cordes, Crawford, Crew, Cruz-Osorio, Cui, {De Laurentis}, Deane, Dempsey, Desvignes, Dexter, Doeleman, Eatough, Farah, Fish, Fomalont, Ford, Fraga-Encinas, Friberg, Fuentes, Galison, Gammie, Garc{\'{i}}a, Gelles, Gentaz, Georgiev, Gold, G{\'{o}}mez-Ruiz, Gu, Gurwell, Hada, Haggard, Hecht, Hesper, Himwich, Ho, Ho, Honma, Huang, Huang, Hughes, Ikeda, Inoue, James, Jannuzi, Jeter, Jiang, Jimenez-Rosales, Jorstad,
  Jung, Karami, Karuppusamy, Kawashima, Keating, Kettenis, Kim, Kim, Kim, Kino, Kofuji, Koyama, Kramer, Kramer, Kuo, Lauer, Lee, Levis, Li, Li, Lindqvist, Lico, Lindahl, Liu, Lo, Lobanov, Loinard, Lonsdale, Lu, MacDonald, Mao, Marchili, Marscher, Mart{\'{i}}-Vidal, Matsushita, Matthews, Medeiros, Menten, Mizuno, Moran, Moriyama, Moscibrodzka, Musoke, Mej{\'{i}}as, Nagai, Nagar, Nakamura, Narayan, Narayanan, Natarajan, Nathanail, Neilsen, Neri, Noutsos, Nowak, Okino, Olivares, Ortiz-Le{\'{o}}n, Oyama, {\"{O}}zel, Palumbo, Park, Patel, Pen, Pi{\'{e}}tu, Plambeck, PopStefanija, Porth, P{\"{o}}tzl, Prather, Preciado-L{\'{o}}pez, Psaltis, Pu, Rao, Rawlings, Raymond, Rezzolla, Ricarte, Ripperda, Rogers, Rose, Roshanineshat, Rottmann, Roy, Ruszczyk, S{\'{a}}nchez, S{\'{a}}nchez-Arguelles, Sasada, Savolainen, Schloerb, Schuster, Shao, Shen, Small, Sohn, SooHoo, Sun, Tazaki, Tetarenko, Tiede, Tilanus, Titus, Torne, Trent, Traianou, Trippe, van Bemmel, van Langevelde, van Rossum, Wagner, Ward-Thompson, Wardle,
  Weintroub, Wex, Wharton, Wong, Wu, Yoon, Young, Young, Younsi, Yuan, Yuan, Zensus, Zhao, \& Zhao}]{janssen2021}
Janssen, M., Falcke, H., Kadler, M., {et~al.} 2021, Nature Astronomy, 5, 1017, \dodoi{10.1038/s41550-021-01417-w}

\bibitem[{{Janssen, M.} {et~al.}(2019){Janssen, M.}, {Goddi, C.}, {van Bemmel, I. M.}, {Kettenis, M.}, {Small, D.}, {Liuzzo, E.}, {Rygl, K.}, {Mart\'{\i}-Vidal, I.}, {Blackburn, L.}, {Wielgus, M.}, \& {Falcke, H.}}]{janssen2019}
{Janssen, M.}, {Goddi, C.}, {van Bemmel, I. M.}, {et~al.} 2019, A\&A, 626, A75, \dodoi{10.1051/0004-6361/201935181}

\bibitem[{{Johnson}(2016)}]{johnson2016}
{Johnson}, M.~D. 2016, \apj, 833, 74, \dodoi{10.3847/1538-4357/833/1/74}

\bibitem[{{Johnson} {et~al.}(2018){Johnson}, {Narayan}, {Psaltis}, {Blackburn}, {Kovalev}, {Gwinn}, {Zhao}, {Bower}, {Moran}, {Kino}, {Kramer}, {Akiyama}, {Dexter}, {Broderick}, \& {Sironi}}]{johnson2018}
{Johnson}, M.~D., {Narayan}, R., {Psaltis}, D., {et~al.} 2018, \apj, 865, 104, \dodoi{10.3847/1538-4357/aadcff}

\bibitem[{{Johnson} {et~al.}(2020){Johnson}, {Lupsasca}, {Strominger}, {Wong}, {Hadar}, {Kapec}, {Narayan}, {Chael}, {Gammie}, {Galison}, {Palumbo}, {Doeleman}, {Blackburn}, {Wielgus}, {Pesce}, {Farah}, \& {Moran}}]{Johnson_2020_photon_ring}
{Johnson}, M.~D., {Lupsasca}, A., {Strominger}, A., {et~al.} 2020, Science Advances, 6, eaaz1310, \dodoi{10.1126/sciadv.aaz1310}

\bibitem[{{Johnson} {et~al.}(2023){Johnson}, {Akiyama}, {Blackburn}, {Bouman}, {Broderick}, {Cardoso}, {Fender}, {Fromm}, {Galison}, {G{\'o}mez}, {Haggard}, {Lister}, {Lobanov}, {Markoff}, {Narayan}, {Natarajan}, {Nichols}, {Pesce}, {Younsi}, {Chael}, {Chatterjee}, {Chaves}, {Doboszewski}, {Dodson}, {Doeleman}, {Elder}, {Fitzpatrick}, {Haworth}, {Houston}, {Issaoun}, {Kovalev}, {Levis}, {Lico}, {Marcoci}, {Martens}, {Nagar}, {Oppenheimer}, {Palumbo}, {Ricarte}, {Rioja}, {Roelofs}, {Thresher}, {Tiede}, {Weintroub}, \& {Wielgus}}]{ngeht_ksg}
{Johnson}, M.~D., {Akiyama}, K., {Blackburn}, L., {et~al.} 2023, Galaxies, 11, 61, \dodoi{10.3390/galaxies11030061}

\bibitem[{{Jorstad} {et~al.}(2023){Jorstad}, {Wielgus}, {Lico}, {Issaoun}, {Broderick}, {Pesce}, {Liu}, {Zhao}, {Krichbaum}, {Blackburn}, {Chan}, {Janssen}, {Ramakrishnan}, {Akiyama}, {Alberdi}, {Algaba}, {Bouman}, {Cho}, {Fuentes}, {G{\'o}mez}, {Gurwell}, {Johnson}, {Kim}, {Lu}, {Mart{\'\i}-Vidal}, {Moscibrodzka}, {P{\"o}tzl}, {Traianou}, {van Bemmel}, {Alef}, {Anantua}, {Asada}, {Azulay}, {Bach}, {Baczko}, {Ball}, {Balokovi{\'c}}, {Barrett}, {Baub{\"o}ck}, {Benson}, {Bintley}, {Blundell}, {Bower}, {Boyce}, {Bremer}, {Brinkerink}, {Brissenden}, {Britzen}, {Broguiere}, {Bronzwaer}, {Bustamante}, {Byun}, {Carlstrom}, {Ceccobello}, {Chael}, {Chatterjee}, {Chatterjee}, {Chen}, {Chen}, {Cheng}, {Christian}, {Conroy}, {Conway}, {Cordes}, {Crawford}, {Crew}, {Cruz-Osorio}, {Cui}, {Davelaar}, {De Laurentis}, {Deane}, {Dempsey}, {Desvignes}, {Dexter}, {Dhruv}, {Doeleman}, {Dougal}, {Dzib}, {Eatough}, {Emami}, {Falcke}, {Farah}, {Fish}, {Fomalont}, {Ford}, {Fraga-Encinas}, {Freeman}, {Friberg}, {Fromm}, {Galison},
  {Gammie}, {Garc{\'\i}a}, {Gentaz}, {Georgiev}, {Goddi}, {Gold}, {G{\'o}mez-Ruiz}, {Gu}, {Hada}, {Haggard}, {Haworth}, {Hecht}, {Hesper}, {Heumann}, {Ho}, {Ho}, {Honma}, {Huang}, {Huang}, {Hughes}, {Ikeda}, {Impellizzeri}, {Inoue}, {James}, {Jannuzi}, {Jeter}, {Jiang}, {Jim{\'e}nez-Rosales}, {Joshi}, {Jung}, {Karami}, {Karuppusamy}, {Kawashima}, {Keating}, {Kettenis}, {Kim}, {Kim}, {Kim}, {Kino}, {Koay}, {Kocherlakota}, {Kofuji}, {Koyama}, {Kramer}, {Kramer}, {Kuo}, {La Bella}, {Lauer}, {Lee}, {Lee}, {Leung}, {Levis}, {Li}, {Lindahl}, {Lindqvist}, {Lisakov}, {Liu}, {Liuzzo}, {Lo}, {Lobanov}, {Loinard}, {Lonsdale}, {MacDonald}, {Mao}, {Marchili}, {Markoff}, {Marrone}, {Marscher}, {Matsushita}, {Matthews}, {Medeiros}, {Menten}, {Michalik}, {Mizuno}, {Mizuno}, {Moran}, {Moriyama}, {M{\"u}ller}, {Mus}, {Musoke}, {Myserlis}, {Nadolski}, {Nagai}, {Nagar}, {Nakamura}, {Narayan}, {Narayanan}, {Natarajan}, {Nathanail}, {Fuentes}, {Neilsen}, {Neri}, {Ni}, {Noutsos}, {Nowak}, {Oh}, {Okino}, {Olivares},
  {Ortiz-Le{\'o}n}, {Oyama}, {{\"O}zel}, {Palumbo}, {Paraschos}, {Park}, {Parsons}, {Patel}, {Pen}, {Pi{\'e}tu}, {Plambeck}, {PopStefanija}, {Porth}, {Prather}, {Preciado-L{\'o}pez}, {Psaltis}, {Pu}, {Rao}, {Rawlings}, {Raymond}, {Rezzolla}, {Ricarte}, {Ripperda}, {Roelofs}, {Rogers}, {Ros}, {Romero-Ca{\~n}izales}, {Roshanineshat}, {Rottmann}, {Roy}, {Ruiz}, {Ruszczyk}, {Rygl}, {S{\'a}nchez}, {S{\'a}nchez-Arg{\"u}elles}, {S{\'a}nchez-Portal}, {Sasada}, {Satapathy}, {Savolainen}, {Schloerb}, {Schonfeld}, {Schuster}, {Shao}, {Shen}, {Small}, {Sohn}, {SooHoo}, {Souccar}, {Sun}, {Tazaki}, {Tetarenko}, {Tiede}, {Tilanus}, {Titus}, {Torne}, {Trent}, {Trippe}, {Turk}, {van Langevelde}, {van Rossum}, {Vos}, {Wagner}, {Ward-Thompson}, {Wardle}, {Weintroub}, {Wex}, {Wharton}, {Wiik}, {Witzel}, {Wondrak}, {Wong}, {Wu}, {Yamaguchi}, {Yoon}, {Young}, {Young}, {Younsi}, {Yuan}, {Yuan}, {Zensus}, {Zhang}, \& {Zhao}}]{Jorstad2023}
{Jorstad}, S., {Wielgus}, M., {Lico}, R., {et~al.} 2023, \apj, 943, 170, \dodoi{10.3847/1538-4357/acaea8}

\bibitem[{{Kerr} {et~al.}(2014){Kerr}, {Pan}, {Claude}, {Dindo}, {Lichtenberger}, {Effland}, \& {Lauria}}]{Kerr_2014}
{Kerr}, A.~R., {Pan}, S.-K., {Claude}, S. M.~X., {et~al.} 2014, IEEE Transactions on Terahertz Science and Technology, 4, 201, \dodoi{10.1109/TTHZ.2014.2302537}

\bibitem[{Kim {et~al.}(2020)Kim, Krichbaum, Broderick, Wielgus, Blackburn, G{\'{o}}mez, Johnson, Bouman, Chael, Akiyama, Jorstad, Marscher, Issaoun, Janssen, Chan, Savolainen, Pesce, {\"{O}}zel, Alberdi, Alef, Asada, Azulay, Baczko, Ball, Balokovi{\'{c}}, Barrett, Bintley, Boland, Bower, Bremer, Brinkerink, Brissenden, Britzen, Broguiere, Bronzwaer, Byun, Carlstrom, Chatterjee, Chatterjee, Chen, Chen, Cho, Christian, Conway, Cordes, Crew, Cui, Davelaar, {De Laurentis}, Deane, Dempsey, Desvignes, Dexter, Doeleman, Eatough, Falcke, Fish, Fomalont, Fraga-Encinas, Friberg, Fromm, Galison, Gammie, Garc{\'{i}}a, Gentaz, Georgiev, Goddi, Gold, G{\'{o}}mez-Ruiz, Gu, Gurwell, Hada, Hecht, Hesper, Ho, Ho, Honma, Huang, Huang, Hughes, Ikeda, Inoue, James, Jannuzi, Jeter, Jiang, Jimenez-Rosales, Jung, Karami, Karuppusamy, Kawashima, Keating, Kettenis, Kim, Kim, Kino, Koay, Koch, Koyama, Kramer, Kramer, Kuo, Lauer, Lee, Li, Li, Lindqvist, Lico, Liu, Liuzzo, Lo, Lobanov, Loinard, Lonsdale, Lu, MacDonald, Mao, Markoff,
  Marrone, Mart{\'{i}}-Vidal, Matsushita, Matthews, Medeiros, Menten, Mizuno, Mizuno, Moran, Moriyama, Moscibrodzka, Musoke, M{\"{u}}ller, Nagai, Nagar, Nakamura, Narayan, Narayanan, Natarajan, Neri, Ni, Noutsos, Okino, Olivares, Ortiz-Le{\'{o}}n, Oyama, Palumbo, Park, Patel, Pen, Pi{\'{e}}tu, Plambeck, PopStefanija, Porth, Prather, Preciado-L{\'{o}}pez, Psaltis, Pu, Ramakrishnan, Rao, Rawlings, Raymond, Rezzolla, Ripperda, Roelofs, Rogers, Ros, Rose, Roshanineshat, Rottmann, Roy, Ruszczyk, Ryan, Rygl, S{\'{a}}nchez, S{\'{a}}nchez-Arguelles, Sasada, Schloerb, Schuster, Shao, Shen, Small, Sohn, SooHoo, Tazaki, Tiede, Tilanus, Titus, Toma, Torne, Trent, Traianou, Trippe, Tsuda, van Bemmel, van Langevelde, van Rossum, Wagner, Wardle, Ward-Thompson, Weintroub, Wex, Wharton, Wong, Wu, Yoon, Young, Young, Younsi, Yuan, Yuan, Zensus, Zhao, Zhao, Zhu, Algaba, Allardi, Amestica, Anczarski, Bach, Baganoff, Beaudoin, Benson, Berthold, Blanchard, Blundell, Bustamente, Cappallo, Castillo-Dom{\'{i}}nguez, Chang, Chang,
  Chang, Chen, Chilson, Chuter, Rosado, Coulson, Crowley, Derome, Dexter, Dornbusch, Dudevoir, Dzib, Eckart, Eckert, Erickson, Everett, Faber, Farah, Fath, Folkers, Forbes, Freund, Gale, Gao, Geertsema, Graham, Greer, Grosslein, Gueth, Haggard, Halverson, Han, Han, Hao, Hasegawa, Henning, Hern{\'{a}}ndez-G{\'{o}}mez, Herrero-Illana, Heyminck, Hirota, Hoge, Huang, {Violette Impellizzeri}, Jiang, John, Kamble, Keisler, Kimura, Kono, Kubo, Kuroda, Lacasse, Laing, Leitch, Li, Lin, Liu, Liu, Lu, Marson, Martin-Cocher, Massingill, Matulonis, McColl, McWhirter, Messias, Meyer-Zhao, Michalik, Monta{\~{n}}a, Montgomerie, Mora-Klein, Muders, Nadolski, Navarro, Neilsen, Nguyen, Nishioka, Norton, Nowak, Nystrom, Ogawa, Oshiro, Oyama, Parsons, Pe{\~{n}}alver, Phillips, Poirier, Pradel, Primiani, Raffin, Rahlin, Reiland, Risacher, Ruiz, S{\'{a}}ez-Mada{\'{i}}n, Sassella, Schellart, Shaw, Silva, Shiokawa, Smith, Snow, Souccar, Sousa, Sridharan, Srinivasan, Stahm, Stark, Story, Timmer, Vertatschitsch, Walther, Wei,
  Whitehorn, Whitney, Woody, Wouterloot, Wright, Yamaguchi, Yu, Zeballos, Zhang, \& Ziurys}]{kim2020}
Kim, J.-y., Krichbaum, T.~P., Broderick, A.~E., {et~al.} 2020, Astronomy \& Astrophysics, 640, A69, \dodoi{10.1051/0004-6361/202037493}

\bibitem[{{Klein} {et~al.}(2014){Klein}, {Ciechanowicz}, {Leinz}, {Heyminck}, {Gusten}, {Kasemann}, {Wunsch}, {Maier}, \& {Sek imoto}}]{klein2014}
{Klein}, T., {Ciechanowicz}, M., {Leinz}, C., {et~al.} 2014, IEEE Transactions on Terahertz Science and Technology, 4, 588, \dodoi{10.1109/TTHZ.2014.2342498}

\bibitem[{Klein {et~al.}(2014)Klein, Ciechanowicz, Leinz, Heyminck, Güsten, Kasemann, Wunsch, Maier, \& Sekimoto}]{Klein_2014}
Klein, T., Ciechanowicz, M., Leinz, C., {et~al.} 2014, IEEE Transactions on Terahertz Science and Technology, 4, 588, \dodoi{10.1109/TTHZ.2014.2342498}

\bibitem[{Koay {et~al.}(2020)Koay, Matsushita, Asada, Patel, Huang, Bower, Chang, Chen, Chen, Chilson, Han, Ho, Huang, Inoue, Jiang, Koch, Koyama, Kuo, Kubo, Li, Liu, Liu, Lo, Martin-Cocher, Nakamura, Norton, Nystrom, Oshiro, Park, Pu, Raffin, Rao, Romero-Canizales, Shaw, Sridharan, Srinivasan, Wei, \& Yu}]{Koay_2020}
Koay, J.~Y., Matsushita, S., Asada, K., {et~al.} 2020, in Ground-based and Airborne Telescopes VIII, ed. H.~K. Marshall, J.~Spyromilio, \& T.~Usuda, Vol. 11445, International Society for Optics and Photonics (SPIE), 114450Q.
\newblock \url{https://doi.org/10.1117/12.2561491}

\bibitem[{{Krichbaum} {et~al.}(1997){Krichbaum}, {Graham}, {Greve}, {Wink}, {Alcolea}, {Colomer}, {de Vicente}, {Baudry}, {Gomez-Gonzalez}, {Grewing}, \& {Witzel}}]{Krichbaum_1997}
{Krichbaum}, T.~P., {Graham}, D.~A., {Greve}, A., {et~al.} 1997, \aap, 323, L17

\bibitem[{{Krichbaum} {et~al.}(1998){Krichbaum}, {Graham}, {Witzel}, {Greve}, {Wink}, {Grewing}, {Colomer}, {de Vicente}, {Gomez-Gonzalez}, {Baudry}, \& {Zensus}}]{Krichbaum_1998}
{Krichbaum}, T.~P., {Graham}, D.~A., {Witzel}, A., {et~al.} 1998, \aap, 335, L106

\bibitem[{{Krichbaum} {et~al.}(2002){Krichbaum}, {Graham}, {Alef}, {Polatidis}, {Bach}, {Witzel}, {Zensus}, {Greve}, {Grewing}, {Doeleman}, {Phillips}, {Rogers}, {Titus}, {Fagg}, {Strittmatter}, {Wilson}, {Ziurys}, {Freund}, {Peltonen}, {Urpo}, {Rantakyr}, {Conway}, \& {Booth}}]{Krichbaum_2002}
{Krichbaum}, T.~P., {Graham}, D.~A., {Alef}, W., {et~al.} 2002, in Proceedings of the 6th EVN Symposium, 125

\bibitem[{Levy {et~al.}(1996)Levy, Antennas, \& Society}]{levy1996structural}
Levy, R., Antennas, I., \& Society, P. 1996, Structural Engineering of Microwave Antennas: For Electrical, Mechanical, and Civil Engineering (IEEE Press).
\newblock \url{https://books.google.com/books?id=qPV_QgAACAAJ}

\bibitem[{{Liebe}(1985)}]{liebe1985}
{Liebe}, H.~J. 1985, Radio Science, 20, 1069, \dodoi{10.1029/RS020i005p01069}

\bibitem[{Lo {et~al.}(2023)Lo, Asada, Matsushita, Pu, Nakamura, Bower, Park, \& Inoue}]{Lo_2023}
Lo, W.-P., Asada, K., Matsushita, S., {et~al.} 2023, The Astrophysical Journal, 950, 10, \dodoi{10.3847/1538-4357/acc855}

\bibitem[{{Lobanov}(1998)}]{lobanov1998}
{Lobanov}, A.~P. 1998, \aap, 330, 79, \dodoi{10.48550/arXiv.astro-ph/9712132}

\bibitem[{{Mahieu} {et~al.}(2012){Mahieu}, {Maier}, {Lazareff}, {Navarrini}, {Celestin}, {Chalain}, {Geoffroy}, {Laslaz}, \& {Perrin}}]{mahieu2012}
{Mahieu}, S., {Maier}, D., {Lazareff}, B., {et~al.} 2012, IEEE Transactions on Terahertz Science and Technology, 2, 29, \dodoi{10.1109/TTHZ.2011.2177734}

\bibitem[{{Maier} {et~al.}(2005){Maier}, {Barbier}, {Lazareff}, \& {Schuster}}]{maier2005}
{Maier}, D., {Barbier}, A., {Lazareff}, B., \& {Schuster}, K.~F. 2005, in Sixteenth International Symposium on Space Terahertz Technology, 428--431

\bibitem[{{Maier} {et~al.}(2012){Maier}, {Reverdy}, {Billon-Pierron}, \& {Barbier}}]{maier2012}
{Maier}, D., {Reverdy}, J., {Billon-Pierron}, D., \& {Barbier}, A. 2012, IEEE Transactions on Terahertz Science and Technology, 2, 215, \dodoi{10.1109/TTHZ.2011.2180609}

\bibitem[{{Mangum} {et~al.}(2006){Mangum}, {Baars}, {Greve}, {Lucas}, {Snel}, {Wallace}, \& {Holdaway}}]{2006PASP..118.1257M}
{Mangum}, J.~G., {Baars}, J. W.~M., {Greve}, A., {et~al.} 2006, \pasp, 118, 1257, \dodoi{10.1086/508298}

\bibitem[{{Mart{\'{\i}}-Vidal} {et~al.}(2016){Mart{\'{\i}}-Vidal}, {Roy}, {Conway}, \& {Zensus}}]{martividal2016}
{Mart{\'{\i}}-Vidal}, I., {Roy}, A., {Conway}, J., \& {Zensus}, A.~J. 2016, \aap, 587, A143, \dodoi{10.1051/0004-6361/201526063}

\bibitem[{{Matsushita} {et~al.}(1999){Matsushita}, {Matsuo}, {Pardo}, \& {Radford}}]{Matsushita_1999}
{Matsushita}, S., {Matsuo}, H., {Pardo}, J.~R., \& {Radford}, S. J.~E. 1999, \pasj, 51, 603, \dodoi{10.1093/pasj/51.5.603}

\bibitem[{Matsushita {et~al.}(2016)Matsushita, Asada, Martin-Cocher, Chen, Ho, Inoue, Koch, Paine, \& Turner}]{Matsushita_2017}
Matsushita, S., Asada, K., Martin-Cocher, P.~L., {et~al.} 2016, Publications of the Astronomical Society of the Pacific, 129, 025001, \dodoi{10.1088/1538-3873/129/972/025001}

\bibitem[{{Matsushita} {et~al.}(2018){Matsushita}, {Asada}, {Inoue}, {Nishioka}, {Huang}, {Patel}, {Koay}, {Koyama}, {Koch}, {Meyer-Zhao}, {Lin}, {Ho}, {Chen}, {Norton}, {Liu}, {Yu}, {Byun}, {Algaba Marcos}, {Allardi}, {Bower}, {Chang}, {Chen}, {Chilson}, {Faber}, {Han}, {Huang}, {Jiang}, {Kubo}, {Liu}, {Lo}, {Martin-Cocher}, {Nakamura}, {Nystrom}, {Oshiro}, {Pu}, {Raffin}, {Shaw}, {Snow}, {Srinivasan}, {Wei}, {Berthold}, {Bintley}, {Dempsey}, {Friberg}, {Walther}, {Weintroub}, {Young}, {Young}, {Sridharan}, {Doeleman}, {Brissenden}, {Ogawa}, {Kimura}, {Hasegawa}, {Hao}, {Han}, {Chang}, \& {Lu}}]{matsushita2018}
{Matsushita}, S., {Asada}, K., {Inoue}, M., {et~al.} 2018, in Society of Photo-Optical Instrumentation Engineers (SPIE) Conference Series, Vol. 10700, Ground-based and Airborne Telescopes VII, ed. H.~K. {Marshall} \& J.~{Spyromilio}, 1070029

\bibitem[{{Matsushita} {et~al.}(2022){Matsushita}, {Martin-Cocher}, {Paine}, {Huang}, {Patel}, {Asada}, {Chen}, {Ho}, {Inoue}, {Koch}, \& {Norton}}]{Matsushita_2022}
{Matsushita}, S., {Martin-Cocher}, P.~L., {Paine}, S.~N., {et~al.} 2022, \pasp, 134, 125002, \dodoi{10.1088/1538-3873/acac51}

\bibitem[{{Matthews} {et~al.}(2018){Matthews}, {Crew}, {Doeleman}, {Lacasse}, {Saez}, {Alef}, {Akiyama}, {Amestica}, {Anderson}, {Barkats}, {Baudry}, {Brogui{\`e}re}, {Escoffier}, {Fish}, {Greenberg}, {Hecht}, {Hiriart}, {Hirota}, {Honma}, {Ho}, {Impellizzeri}, {Inoue}, {Kohno}, {Lopez}, {Mart{\'\i}-Vidal}, {Messias}, {Meyer-Zhao}, {Mora-Klein}, {Nagar}, {Nishioka}, {Oyama}, {Pankratius}, {Perez}, {Phillips}, {Pradel}, {Rottmann}, {Roy}, {Ruszczyk}, {Shillue}, {Suzuki}, \& {Treacy}}]{matthews2018}
{Matthews}, L.~D., {Crew}, G.~B., {Doeleman}, S.~S., {et~al.} 2018, \pasp, 130, 015002, \dodoi{10.1088/1538-3873/aa9c3d}

\bibitem[{{Meledin, D.} {et~al.}(2022){Meledin, D.}, {Lapkin, I.}, {Fredrixon, M.}, {Sundin, E.}, {Ferm, S.-E.}, {Pavolotsky, A.}, {Strandberg, M.}, {Desmaris, V.}, {López, C.}, {Bergman, P.}, {Olberg, M.}, {Conway, J.}, {Torstensson, K.}, {Durán, C.}, {Montenegro-Montes, F. M.}, {De Breuck, C.}, \& {Belitsky, V.}}]{meledin_2022}
{Meledin, D.}, {Lapkin, I.}, {Fredrixon, M.}, {et~al.} 2022, A\&A, 668, A2, \dodoi{10.1051/0004-6361/202244211}

\bibitem[{{Okino} {et~al.}(2022){Okino}, {Akiyama}, {Asada}, {G{\'o}mez}, {Hada}, {Honma}, {Krichbaum}, {Kino}, {Nagai}, {Bach}, {Blackburn}, {Bouman}, {Chael}, {Crew}, {Doeleman}, {Fish}, {Goddi}, {Issaoun}, {Johnson}, {Jorstad}, {Koyama}, {Lonsdale}, {Lu}, {Mart{\'\i}-Vidal}, {Matthews}, {Mizuno}, {Moriyama}, {Nakamura}, {Pu}, {Ros}, {Savolainen}, {Tazaki}, {Wagner}, {Wielgus}, \& {Zensus}}]{okino2022}
{Okino}, H., {Akiyama}, K., {Asada}, K., {et~al.} 2022, The Astrophysical Journal, 940, 65, \dodoi{10.3847/1538-4357/ac97e5}

\bibitem[{{Padin} {et~al.}(1990){Padin}, {Woody}, {Hodges}, {Rogers}, {Emerson}, {Jewell}, {Lamb}, {Perfetto}, \& {Wright}}]{padin1990}
{Padin}, S., {Woody}, D.~P., {Hodges}, M.~W., {et~al.} 1990, \apjl, 360, L11, \dodoi{10.1086/185800}

\bibitem[{Paine(2022)}]{paine2022}
Paine, S. 2022, \dodoi{10.5281/zenodo.6774378}

\bibitem[{{Palumbo} {et~al.}(2023){Palumbo}, {Wong}, {Chael}, \& {Johnson}}]{Palumbo_2023}
{Palumbo}, D. C.~M., {Wong}, G.~N., {Chael}, A., \& {Johnson}, M.~D. 2023, \apjl, 952, L31, \dodoi{10.3847/2041-8213/ace630}

\bibitem[{{Paraschos} {et~al.}(2024){Paraschos}, {Kim}, {Wielgus}, {R{\"o}der}, {Krichbaum}, {Ros}, {Agudo}, {Myserlis}, {Moscibrodzka}, {Traianou}, {Zensus}, {Blackburn}, {Chan}, {Issaoun}, {Janssen}, {Johnson}, {Fish}, \& {EHTC}}]{Paraschos2024}
{Paraschos}, G.~F., {Kim}, J.~Y., {Wielgus}, M., {et~al.} 2024, \aap, 682, L3, \dodoi{10.1051/0004-6361/202348308}

\bibitem[{{Pesce} {et~al.}(2021){Pesce}, {Palumbo}, {Narayan}, {Blackburn}, {Doeleman}, {Johnson}, {Ma}, {Nagar}, {Natarajan}, \& {Ricarte}}]{Pesce_2021}
{Pesce}, D.~W., {Palumbo}, D. C.~M., {Narayan}, R., {et~al.} 2021, \apj, 923, 260, \dodoi{10.3847/1538-4357/ac2eb5}

\bibitem[{{Primiani} {et~al.}(2016){Primiani}, {Young}, {Young}, {Patel}, {Wilson}, {Vertatschitsch}, {Chitwood}, {Srinivasan}, {MacMahon}, \& {Weintroub}}]{primiani2016}
{Primiani}, R.~A., {Young}, K.~H., {Young}, A., {et~al.} 2016, Journal of Astronomical Instrumentation, 5, 1641006, \dodoi{10.1142/S2251171716410063}

\bibitem[{{Raffin} {et~al.}(2016){Raffin}, {Ho}, {Asada}, {Blundell}, {Bower}, {Burgos}, {Chang}, {Chen}, {Christensen}, {Chu}, {Grimes}, {Han}, {Huang}, {Huang}, {Hsieh}, {Inoue}, {Koch}, {Kubo}, {Leiker}, {Lin}, {Liu}, {Lo}, {Martin-Cocher}, {Matsushita}, {Nakamura}, {Meyer-Zhao}, {Nishioka}, {Norton}, {Nystrom}, {Paine}, {Patel}, {Pu}, {Snow}, {Sridharan}, {Srinivasan}, \& {Wang}}]{raffin2016}
{Raffin}, P., {Ho}, P. T.~P., {Asada}, K., {et~al.} 2016, in Society of Photo-Optical Instrumentation Engineers (SPIE) Conference Series, Vol. 9906, Ground-based and Airborne Telescopes VI, ed. H.~J. {Hall}, R.~{Gilmozzi}, \& H.~K. {Marshall}, 99060U

\bibitem[{{Ramakrishnan} {et~al.}(2023){Ramakrishnan}, {Nagar}, {Arratia}, {Hern{\'a}ndez-Y{\'e}venes}, {Pesce}, {Nair}, {Bandyopadhyay}, {Medina-Porcile}, {Krichbaum}, {Doeleman}, {Ricarte}, {Fish}, {Blackburn}, {Falcke}, {Bower}, \& {Natarajan}}]{Venki_2023}
{Ramakrishnan}, V., {Nagar}, N., {Arratia}, V., {et~al.} 2023, Galaxies, 11, 15, \dodoi{10.3390/galaxies11010015}

\bibitem[{Raymond {et~al.}(2021)Raymond, Palumbo, Paine, Blackburn, {C{\'{o}}rdova Rosado}, Doeleman, Farah, Johnson, Roelofs, Tilanus, \& Weintroub}]{raymond2021}
Raymond, A.~W., Palumbo, D., Paine, S.~N., {et~al.} 2021, The Astrophysical Journal Supplement Series, 253, 5, \dodoi{10.3847/1538-3881/abc3c3}

\bibitem[{{Readhead} {et~al.}(1983){Readhead}, {Mason}, {Mofett}, {Pearson}, {Seielstad}, {Woody}, {Backer}, {Plambeck}, {Welch}, {Wright}, {Rogers}, {Webber}, {Shapiro}, {Moran}, {Goldsmith}, {Predmore}, {Baath}, \& {Ronnang}}]{readhead1983}
{Readhead}, A.~C.~S., {Mason}, C.~R., {Mofett}, A.~T., {et~al.} 1983, \nat, 303, 504, \dodoi{10.1038/303504a0}

\bibitem[{{Rioja} {et~al.}(2023){Rioja}, {Dodson}, \& {Asaki}}]{ngeht_fpt}
{Rioja}, M.~J., {Dodson}, R., \& {Asaki}, Y. 2023, Galaxies, 11, 16, \dodoi{10.3390/galaxies11010016}

\bibitem[{Rogers \& Moran(1981)}]{rogers1981}
Rogers, A.~E., \& Moran, J.~M. 1981, IEEE Transactions on Instrumentation and Measurement, IM-30, 283, \dodoi{10.1109/TIM.1981.6312409}

\bibitem[{{Rogers} {et~al.}(1995){Rogers}, {Doeleman}, \& {Moran}}]{rogers1995}
{Rogers}, A. E.~E., {Doeleman}, S.~S., \& {Moran}, J.~M. 1995, \aj, 109, 1391, \dodoi{10.1086/117371}

\bibitem[{Rogers {et~al.}(1984)Rogers, Moffet, Backer, \& Moran}]{rogers1984}
Rogers, A. E.~E., Moffet, A.~T., Backer, D.~C., \& Moran, J.~M. 1984, Radio Science, 19, 1552, \dodoi{10.1029/RS019i006p01552}

\bibitem[{{Thompson} {et~al.}(2017){Thompson}, {Moran}, \& {Swenson}}]{TMS2017}
{Thompson}, A.~R., {Moran}, J.~M., \& {Swenson}, Jr., G.~W. 2017, {Interferometry and Synthesis in Radio Astronomy, 3rd Edition}, \dodoi{10.1007/978-3-319-44431-4}

\bibitem[{{Tong} {et~al.}(2005){Tong}, {Blundell}, {Megerian}, {Stern}, {Pan}, \& {Pospieszalski}}]{tong2005}
{Tong}, C. Y.~E., {Blundell}, R., {Megerian}, K.~G., {et~al.} 2005, IEEE Transactions on Applied Superconductivity, 15, 490, \dodoi{10.1109/TASC.2005.849885}

\bibitem[{{Tong} {et~al.}(2013){Tong}, {Grimes}, {Blundell}, {Wang}, \& {Noguchi}}]{Tong_2013}
{Tong}, C.-Y.~E., {Grimes}, P., {Blundell}, R., {Wang}, M.-J., \& {Noguchi}, T. 2013, IEEE Transactions on Terahertz Science and Technology, 3, 428, \dodoi{10.1109/TTHZ.2013.2259624}

\bibitem[{Treuhaft \& Lanyi(1987)}]{treuhaft1987}
Treuhaft, R.~N., \& Lanyi, G.~E. 1987, Radio Science, 22, 251, \dodoi{https://doi.org/10.1029/RS022i002p00251}

\bibitem[{{Tuccari} {et~al.}(2017){Tuccari}, {Alef}, {Wunderlich}, {Buttaccio}, {Graham}, {Rottmann}, {Bertarini}, {Roy}, {Dornbusch}, {Felke}, {Casey}, \& {Lindqvist}}]{Tuccari2017}
{Tuccari}, G., {Alef}, W., {Wunderlich}, M., {et~al.} 2017, in 23rd European VLBI Group for Geodesy and Astrometry Working Meeting, ed. R.~{Haas} \& G.~{Elgered}, 78--80

\bibitem[{{Vertatschitsch} {et~al.}(2015){Vertatschitsch}, {Primiani}, {Young}, {Weintroub}, {Crew}, {McWhirter}, {Beaudoin}, {Doeleman}, \& {Blackburn}}]{Vertatschitsch2015}
{Vertatschitsch}, L., {Primiani}, R., {Young}, A., {et~al.} 2015, \pasp, 127, 1226, \dodoi{10.1086/684513}

\bibitem[{Whitney {et~al.}(2004)Whitney, Cappallo, Aldrich, Anderson, Bos, Casse, Goodman, Parsley, Pogrebenko, Schilizzi, \& Smythe}]{whitney2004}
Whitney, A.~R., Cappallo, R., Aldrich, W., {et~al.} 2004, Radio Science, 39, \dodoi{10.1029/2002RS002820}

\bibitem[{{Wielgus} {et~al.}(2024){Wielgus}, {Issaoun}, {Mart{\'\i}-Vidal}, {Emami}, {Moscibrodzka}, {Brinkerink}, {Goddi}, \& {Fomalont}}]{Wielgus_scattering_screen_2023}
{Wielgus}, M., {Issaoun}, S., {Mart{\'\i}-Vidal}, I., {et~al.} 2024, \aap, 682, A97, \dodoi{10.1051/0004-6361/202347772}

\bibitem[{{Young} {et~al.}(2016){Young}, {Primiani}, {Weintroub}, {Moran}, {Young}, {Blackburn}, {Johnson}, \& {Wilson}}]{young2016}
{Young}, A., {Primiani}, R., {Weintroub}, J., {et~al.} 2016, in 2016 IEEE International Symposium on Phased Array Systems and Technology (PAST), 1--8

\bibitem[{{Zhao} {et~al.}(2022){Zhao}, {G{\'o}mez}, {Fuentes}, {Krichbaum}, {Traianou}, {Lico}, {Cho}, {Ros}, {Komossa}, {Akiyama}, {Asada}, {Blackburn}, {Britzen}, {Bruni}, {Crew}, {Dahale}, {Dey}, {Gold}, {Gopakumar}, {Issaoun}, {Janssen}, {Jorstad}, {Kim}, {Koay}, {Kovalev}, {Koyama}, {Lobanov}, {Loinard}, {Lu}, {Markoff}, {Marscher}, {Mart{\'\i}-Vidal}, {Mizuno}, {Park}, {Savolainen}, \& {Toscano}}]{zhao2022}
{Zhao}, G.-Y., {G{\'o}mez}, J.~L., {Fuentes}, A., {et~al.} 2022, The Astrophysical Journal, 932, 72, \dodoi{10.3847/1538-4357/ac6b9c}

\end{thebibliography}

\end{document}